\RequirePackage{fix-cm}
\RequirePackage{rotating}
\documentclass[smallextended]{svjour3}       
\smartqed  
\usepackage[utf8]{inputenc}
\usepackage[T1]{fontenc}
\usepackage{graphicx}
\usepackage{caption}
\usepackage{wrapfig, blindtext}
\usepackage{float}
\usepackage{tabularx}
\usepackage{array}
\usepackage{comment}
\usepackage{rotating}
\usepackage{geometry}
\usepackage{caption}
\usepackage{adjustbox}
\usepackage[table,dvipsnames]{xcolor}
\usepackage{booktabs}
\usepackage{multirow, makecell}
\renewcommand\theadfont{\normalsize}

\begin{document}
\title{Semi-Supervised Deep Learning for Multi-Tissue Segmentation from Multi-Contrast MRI}
%
%
%
%
%



\author{Syed Muhammad Anwar \and
        Ismail Irmakci \and Drew A. Torigian \and Sachin Jambawalikar \and Georgios Z. Papadakis \and Can Akgun \and Jutta Ellermann \and Mehmet Akcakaya  \and Ulas Bagci
}

\institute{Dr. Anwar and Dr. Bagci are with the Center for Research in Computer Vision, University of Central Florida, Orlando, Florida, USA. Dr. Irmakci is with Ege University, Izmir, Turkey. Dr. Torigian is with Department of Radiology, Hospital of the University of Pennsylvania, Pennsylvania, USA. Dr. Jambawalikar is with Columbia University, NYC. Dr. Papadakis is with Foundation for Research and Technology Hellas, Crete, Greece. Dr. Akgun is with FlyWheel Inc., Boston, MA. Dr. Akcakaya and Dr. Ellerman are with University of Minnesota, Minniapolis. Corresponding author's email: \email{bagci@ucf.edu}  }

            
              

\date{Received: date / Accepted: date}


\maketitle 

\begin{abstract}
Segmentation of thigh tissues (muscle, fat, inter-muscular adipose tissue (IMAT), bone, and bone marrow) from magnetic resonance imaging (MRI) scans is useful for clinical and research investigations in various conditions  such as  aging, diabetes mellitus, obesity, metabolic syndrome, and their associated comorbidities. Towards a fully automated, robust, and precise quantification of thigh tissues, herein we designed a novel semi-supervised segmentation algorithm based on deep  network architectures. Built upon \textit{Tiramisu} segmentation engine, our proposed deep networks use variational and specially designed targeted dropouts for faster and robust convergence, and utilize multi-contrast MRI scans as input data. In our experiments, we have used $150$ scans from $50$ distinct subjects from the Baltimore Longitudinal Study of Aging (BLSA).  
The proposed system made use of both labeled and unlabeled data with high efficacy for training, and outperformed the current state-of-the-art methods. In particular, dice scores of $97.52\%$, $94.61\%$, $80.14\%$, $95.93\%$, and $96.83\%$ are achieved for muscle, fat, IMAT, bone, and bone marrow segmentation, respectively. Our results indicate that the proposed system can be useful for clinical research studies where volumetric and distributional tissue quantification is pivotal and labeling is a significant issue. To the best of our knowledge, the proposed system is the first attempt at multi-tissue segmentation using a single end-to-end semi-supervised deep learning framework for multi-contrast thigh MRI scans.
\end{abstract}

\section{Introduction}


The body composition of tissues changes over time and human muscles tend to lose strength. This could be due to aging or clinical conditions such as diabetes milletus and metabolic syndrome~\cite{goodpaster2003association}. The muscles and bones in human body are mechano-responsive tissues, whose strength reduces with time due to the accumulation of adipose (fat) tissue. Fat accumulation in bone occurs in the marrow region. When it occurs within the muscle, it is called fat infiltration \cite{10.3945/ajcn.2009.28047}. Fat infiltration in muscles acts as an indicator of various clinical outcomes~\cite{hamrick2016fatty}, although, inter- and intra-muscular fat accumulation is still not well understood. One of the reasons for this research gap is the challenge in precise quantification of fat accumulation in the muscular tissue. An automated system that can effectively segment various tissues within the thigh region could lead to a better understanding of this phenomenon.    

\begin{figure}[t]
\centering
\begin{tabular}{cc}
\multicolumn{1}{c}{\includegraphics[height=2cm,width=4cm]{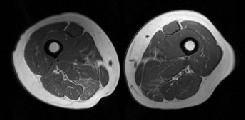}} & 

\multicolumn{1}{c}{\includegraphics[height=2cm,width=4cm]{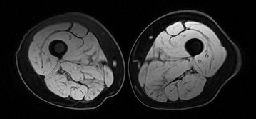}}  \\

\multicolumn{1}{c}{(a) water-and-fat} & \multicolumn{1}{c}{(b) fat-suppressed} \\

\multicolumn{1}{c}{\includegraphics[height=2cm,width=4cm]{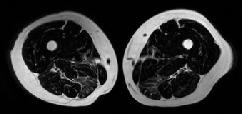}} &

\multicolumn{1}{l}{\includegraphics[height=2cm,width=4cm]{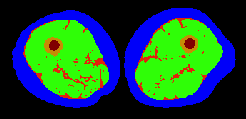}} \\

\multicolumn{1}{c}{(c) water-suppressed} & \multicolumn{1}{c}{(d) Ground Truth (GT)}
\end{tabular}
\caption{An example of a multi-contrast MRI scan (a, b, c) and ground truth tissue labels (d). Fat: blue, muscle: green, bone: orange, bone-marrow: brown, IMAT: red.\label{fig:example_scans}}
\end{figure}

There are numerous skeletal muscles within the human body which help in maintaining posture and making movements. These tissues also contribute to thermo-regulation by keeping an energy balance within the body \cite{porter1995aging}. Among skeletal muscles, the thigh muscle group is one of the largest and plays a very important role in movement and defining human gait. It can be severely effected by a lack of use or aging. Muscular dystrophy is an important condition that needs to be detected at the right time to avoid muscular loss. Fat fraction can be used as a bio-marker to study such kinds of musculo-skeletal diseases \cite{loughran2015improving}, where an accurate segmentation of the muscle tissue becomes an important task. Multi-contrast Magnetic Resonance (MR) images (e.g., water-and-fat, fat-suppressed, and water-suppressed images) as shown in Figure \ref{fig:example_scans} are used to address tissue volume quantification. Since tissue boundaries are often ambiguous, multi-contrasts are often used.  The contrast levels between muscle and subcutaneous fat are difficult to identify with both automated methods and expert manual assessment \cite{gadermayr2018comprehensive}. In this regard, the existing auto-segmentation methods fall short especially when dealing with Inter-Muscular Adipose Tissue (IMAT). Non-invasive imaging techniques are often sought to evaluate fat infiltration and to quantify its extent. However, an automated and accurate image segmentation technique is still required. Although Magnetic Resonance Imaging (MRI) provides good contrast when imaging muscular regions \cite{morrow2016mri}, accurately segmenting pathological tissue using automated methods is challenging. Mostly in current clinical settings, manual segmentation methods are used, but these are tedious and time consuming with a high amount of inter- and intra-observer variability \cite{mhuiris2016reliability}.

Although current efforts are mostly focused on detecting and quantifying IMAT, there is a strong need for quantifying other tissues in the thigh region including, muscle, fat, bone, and bone marrow. In particular, there is still a lack in systematic understanding of how change in various tissue volumes and distribution (across groups or longitudinally) contribute to metabolic health. In this regard, MRI is a frequent imaging choice for the thigh region due to its lack of ionizing radiation, and superior soft tissue contrast compared with other non-invasive imaging methods \cite{diaz2015muscle}. Furthermore, with MRI, multiple contrasts can be applied to the same tissues for enhancing their visualization and analysis. An example of this phenomena is illustrated for the thigh MRI shown in Figure \ref{fig:example_scans}, where three different MRI contrasts are visualized: Water-and-fat, fat-suppressed, and water-suppressed. Despite the provision of complementary tissue information using multi-contrast MRI, current segmentation methods fall short in capturing tissue details especially when IMAT distributions change with respect to varying clinical conditions. Hence, there remains a challenge to develop an automated system that can perform multi-tissue segmentation in multi-contrast thigh MR images with sparsely annotated data.


\subsection{\textbf{Related Works}} 

Deep learning has become the standard choice for addressing various problems in medical image analysis \cite{anwar2018medical}\cite{hussain2018segmentation}\cite{hussein2019lung}, but has not been widely used in musculo-skeletal MRI studies yet. There are various reasons for this including: the time required to generate manual annotations for supervised deep learning, the variability in MRI data, the anatomical differences between subjects, and the large amount of imaging data required for pixel level labeling. A notable study utilizing deep learning algorithms was proposed \cite{gadermayr2019domain}, in which fatty infiltration was evaluated with automatic segmentation using MRIs of healthy subjects. A standard U-Net architecture \cite{ronneberger2015u} with Generative Adversarial Networks (GANs) (to overcome the lack of large scale training data) was utilized to augment texture and shape into segmentation procedure, achieving a dice score of $88\%$. A multi-parametric deep learning network was proposed for generating the tissue map in the thigh region using whole body MRI scans \cite{bocchieri2019multiparametric}. The study used a small cohort of $24$ patients and reported dice scores of $81\%$ and $82\%$ for fat and muscle tissues, respectively. 

Surprisingly, the literature on automatic quantification of thigh tissues from the pre-deep learning era is limited too, where existing studies are focused on IMAT research mostly, and less on multi-tissue quantification with multi-contrast MRI. Different techniques were used for the segmentation of fat and muscle using MR scans from $37$ patients \cite{gadermayr2018comprehensive}, including thresholding, local, and global energy minimization (via graph cut). It was found that graph cut methods gave the best performance for severe fat infiltration where shape prior knowledge was used. A Gaussian mixture model based method was used for segmenting the muscle, fat, and IMAT \cite{makrogiannis2016image}. The clustering based techniques worked well for IMAT, but failed to detect the subcutaneous fat. A novel fuzzy connectivity based segmentation method was used for fat and muscle segmentation in thigh images \cite{irmakci2018novel}. The method used affinity propagation clustering technique to minimize user intervention in the segmentation process while achieving the-state-of-the-art dice score of $84\%$ and $87\%$ for fat and muscle tissues, respectively. A comprehensive evaluation of other pre-deep learning era methods can be found here \cite{irmakci2018novel}. 

The methods presented above either lack in performance or cannot provide an end-to-end solution for multi-tissue segmentation of thigh MR images using a unified framework. Our proposed method uses a single framework in both supervised and semi-supervised settings to segment multi-contrast MR images and achieves a fast and reliable segmentation map for multiple tissues within the thigh region. 


\subsection{\textbf{Our Contributions}} 

We propose a new deep learning framework to accurately segment multiple tissues pertaining to the thigh region using multi-contrast MRI. By employing both supervised and semi-supervised strategies on the Baltimore Longitudinal Study for Aging (BLSA) imaging data, we obtained new state-of-the-art results in quantifying muscle, fat, IMAT, bone, and bone marrow. Our study is novel because no previous study has investigated deep learning as an accurate and efficient solution for multi-tissue quantification from multi-contrast thigh MRI. Furthermore, we incorporated architectural novelties into the deep segmentation problem, where we handled the lack of large amounts of precisely labeled samples using the semi-supervised Tiramisu network. In the last part of the study, we studied the relationship of efficiency and robustness in our proposed semi-supervised deep network. Despite the importance of these concepts (efficiency and robustness), there has been no controlled deep learning based segmentation studies which compare robustness and convergence of the segmentation process when imaging data and ground truth labels are scarce. This is particularly important in the medical imaging field. To address this research gap, we revisited the \textit{dropout} concept in deep learning, and with the help of \textit{Targeted Dropout (TD)} and \textit{Variational Dropout (VD)} anchored into our network, improved segmentation results with faster convergence and better training experience were achieved. Our proposed strategies outperformed the state-of-the-art methods in the literature both in supervised and semi-supervised multi-tissue segmentation problems. We believe that our end-to-end deep network solution will be a significant milestone towards understanding how the inter- and intra-muscular accumulation of adipose tissue affects muscles in particular and metabolic health in general, and would allow us to address muscular dystrophies and the effects of aging.

\section{Methods}\label{meth}
We proposed a semi-supervised approach to segment multiple tissues from three-different contrast MRI scans (Figure~\ref{fig:example_scans} a-c). We also test the applicability of our proposed system under fully supervised setting (to compare results) for tissues whose ground truth annotations are available. In the supervised approach, fat, muscle, bone, and bone-marrow tissues are learned. In our proposed semi-supervised approach, a more challenging tissue IMAT as well as muscle, fat, bone, and bone marrow tissues were learned. 
The proposed methodology is shown in Figure \ref{fig:my_label}, where both supervised and semi-supervised settings are elaborated. After all (three) contrasts of MRI scans (Figure~\ref{fig:example_scans}) were pre-processed to remove inhomogeneity, minimize noise, and standardize intensity scale across subjects, an encoder-decoder style deep network (called \textit{Tiramisu} due to its 103-layers with dense connections) was designed for the segmentation of the thigh region by employing both supervised and semi-supervised techniques. The output of the end-to-end framework is the segmentation map for muscle, fat, IMAT, bone, and bone marrow tissues. 

\subsection{\textbf{Dataset}}

A total of $150$ scans from $50$ subjects were used in our evaluations (BLSA study participants, 3T Philips Achieva MR scanner equipped with a Q-body radio frequency coil for transmission and reception) \cite{56caf3c85e12457eaa97fc0befaa71b8}. The dataset is publicly available once a signed agreement is made with BLSA. Three different T1-weighted MR contrasts (water and fat, fat-suppressed, and water-suppressed) were used, where water and fat suppression were achieved using spectral pre-saturation with inversion recovery (SPIR). These separate image volumes were obtained using a spoiled gradient recalled echo (SPGR) sequence, with coverage from the proximal to distal ends of the femur using $80$ slices, a field of view (FOV) of $440 \times 440 mm^2$, a voxel size of $1 \times 1 mm^2$ in-plane, and slice thickness varying from $1$ mm to $3$ mm in different scans. Acquisition parameters included: Repetition time (TR) = $7.7$ ms, echo time (TE) = $2.4$ ms, number of signal averages (NSA) = $2$, flip angle (FA) = $25$ degrees, and bandwidth of $452$ Hz/pixel. The first image volume (among those three volumes obtained) containing water and fat signal intensities was obtained with the T1 SPGR/fast field echo (FFE) sequence described above. For fat-only and water-only images, water and fat suppression were obtained using spectral pre-saturation with inversion recovery. The age of subjects ranged between 44-89 years ($71 \pm 11$) and the body mass index (BMI) ranged from 18.67 - 45.68 ($26.86 \pm 5.02$). The details of the subjects included in the study are given in Table \ref{table:cohort}. 

\begin {table}[!h]
\centering
\caption{Details about the experimental cohort used in the experiments including subject age (years), weight (kg), height (cm) and Body Mass Index ($Kg/m^2$).}
\label{table:cohort}
\begin{tabularx}{0.8\columnwidth}{@{\extracolsep\fill}lll}
\toprule
\textbf{Characteristics} & \textbf{Range} & \textbf{Mean $\pm$ Std Dev} \\
\midrule

\midrule
\textbf{Age} & [44, 89] years & 71$\pm$11 years \\
\textbf{Weight} & [47.7, 121] kg &  77.5$\pm$17 kg \\
\textbf{Height} & [148, 187.1] cm & 169.32$\pm$9.26 cm \\
\textbf{BMI} & [18.67, 45.68] $kg/m^2$&  26.87$\pm$5.02 $kg/m^2$ \\
\bottomrule
\end{tabularx}

\end{table}
\subsection{\noindent\textbf{Pre-processing MRIs}} 
MR images are affected by various types of artifacts including inhomogeneity (field bias), intensity non-standardness, and inherent noises from the acquisition process \cite{lugauer2015robust}. The removal of bias generates additional noise, and hence a de-noising filter should be applied after bias correction. We used a non-uniform non-parametric intensity normalization technique (N4ITK) \cite{n4itk04} to remove field bias, which itself generated additional noise. An edge-preserving diffusive filter was used for removing this noise, while preserving structures and image sharpness \cite{aurich1995non}. We also standardized the MR image intensities between a minimum and maximum value of $1$ and $4095$, by adopting the MRI scale standardization approach \cite{nyul2000new}.\\

\renewcommand\theadfont{}

\begin{figure}[!t]
\centering
    \includegraphics[width=120mm]{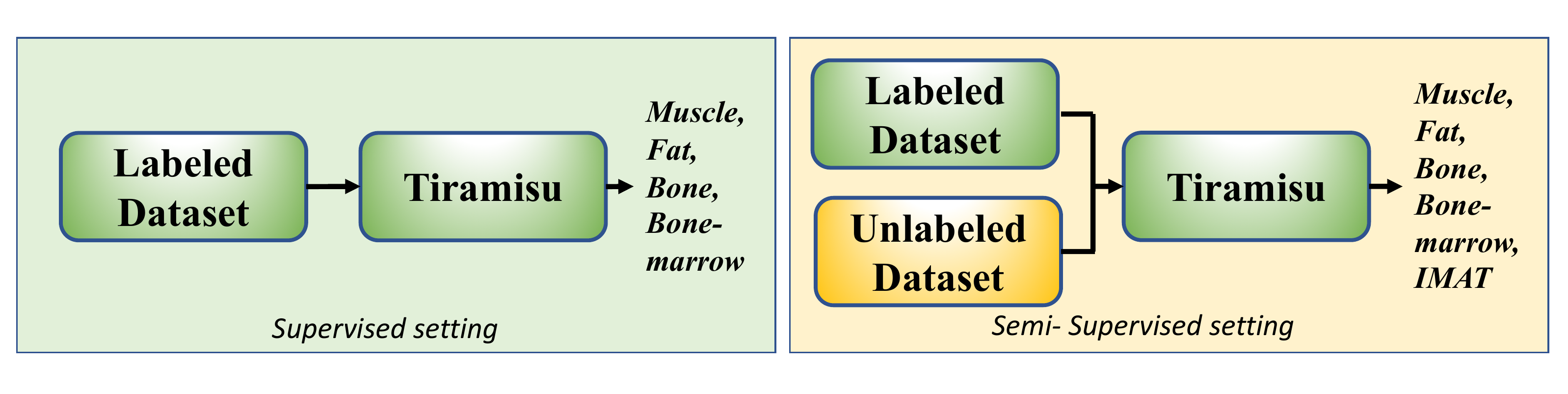}
    \caption{Proposed supervised and semi-supervised segmentation settings. \label{fig:my_label}}
\end{figure}  

\begin{figure}[t]
\centering
\begin{tabular}{cc }
\includegraphics[width=0.75\textwidth]{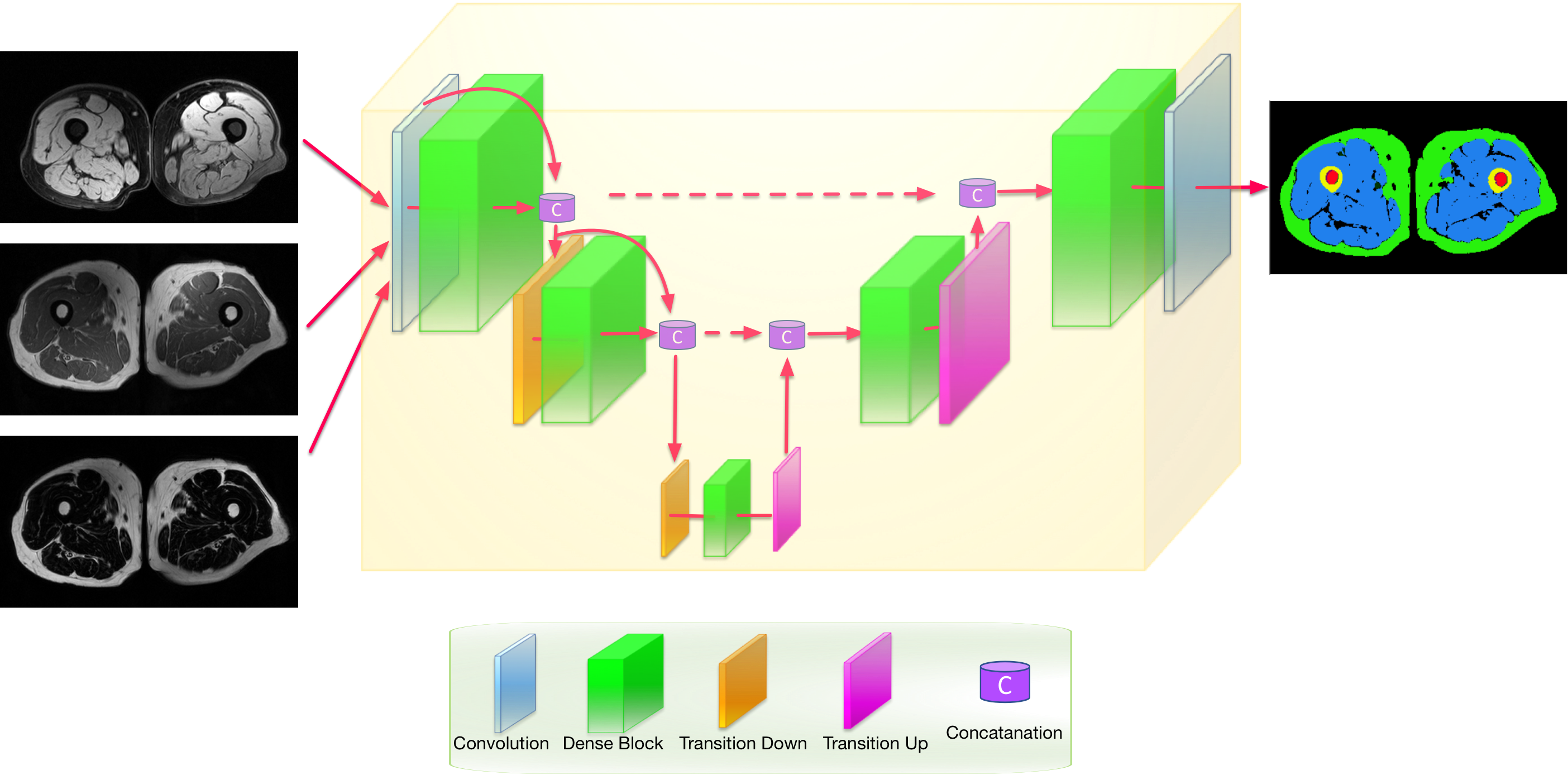} & 	\includegraphics[width=0.25\textwidth]{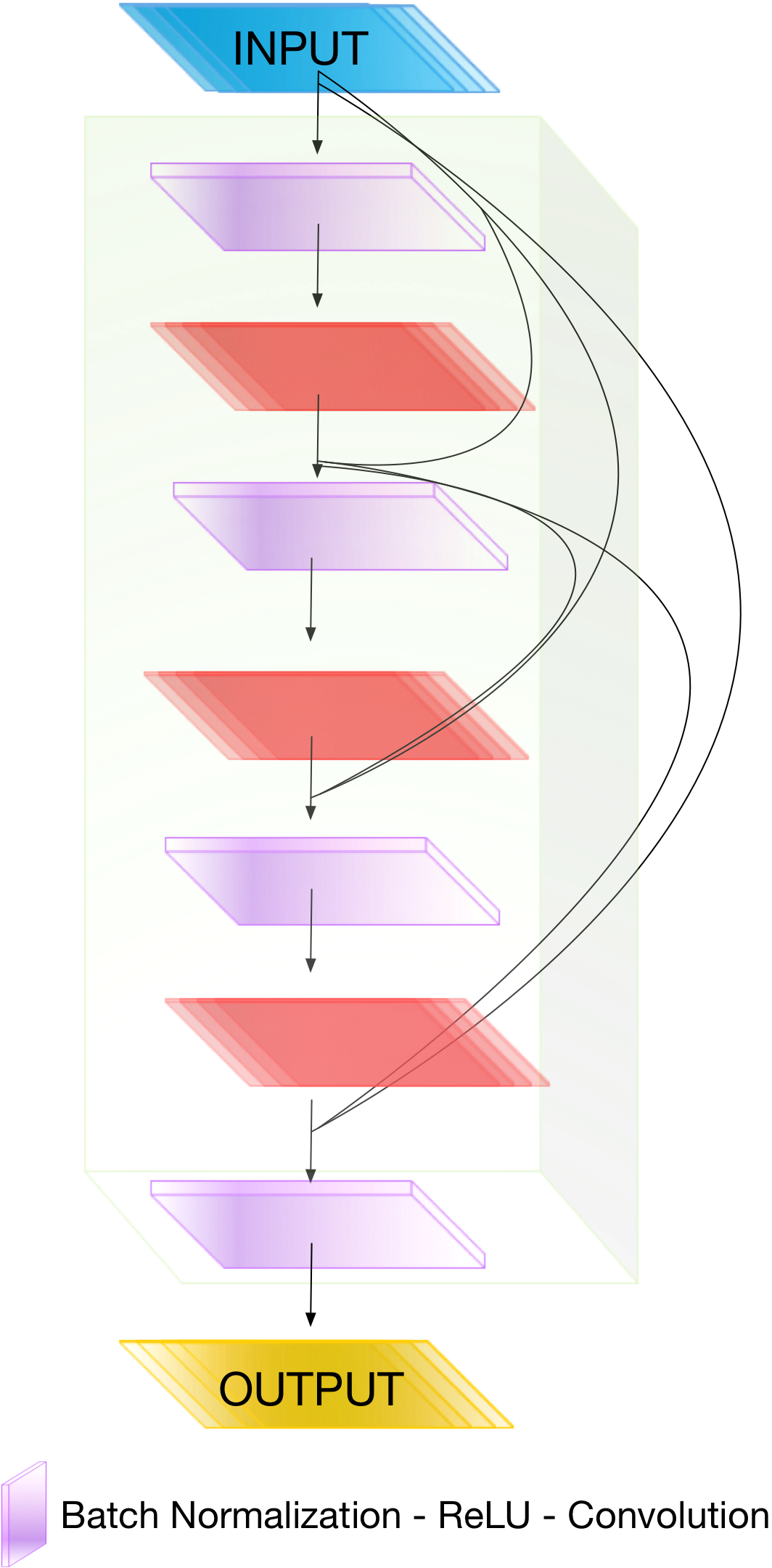}\\
(a) \thead{Tiramisu adopted with multi-contrast MRI \\ and various dropouts.} & (b) \thead{Content of a\\dense block.} \\
\end{tabular}
\caption{Tiramisu architecture with multi-contrast MR as input, and segmentation map as output. Dropouts are embedded within dense blocks. \label{fig:denseNetArchitecture}}
\end{figure}
\subsection{\noindent\textbf{Supervised Segmentation}} 

We created our proposed architecture based on a fully supervised nature before converting it into a semi-supervised framework. The amount of labeled data plays a significant role in the performance of the supervised approach (Figure~\ref{fig:my_label}). We addressed the lack of annotations (labels) for IMAT tissue via a semi-supervised approach. The segmentation of the pre-processed thigh MR images was performed using the densely-connected deep network architecture Tiramisu~\cite{tiramisu} (Figure~\ref{fig:denseNetArchitecture}). Tiramisu follows an encoder-decoder structure and it was originally proposed for semantic segmentation of natural images. 
The network architecture was 
composed of convolution layers, dense blocks, and skip connections which were arranged in a down-sampling/up-sampling pipeline. In the dense block (Figure \ref{fig:denseNetArchitecture} (b)), the information was enhanced by concatenation of layers, where each layer was composed of batch-normalization, ReLU activation function, and convolution. Furthermore, three variations of the Tiramisu network were designed: Tiramisu+R, Tiramisu+VD, and Tiramisu+TD, where R, VD, and TD represent the regular, variational, and targeted dropout, respectively. Four classes of tissues were segmented using the supervised setting including muscle, fat, bone, and bone-marrow. \\   

\subsection{\noindent\textbf{Semi-supervised Segmentation}} 

\begin{figure}[!t]
    \centering
    \includegraphics[width = 140mm]{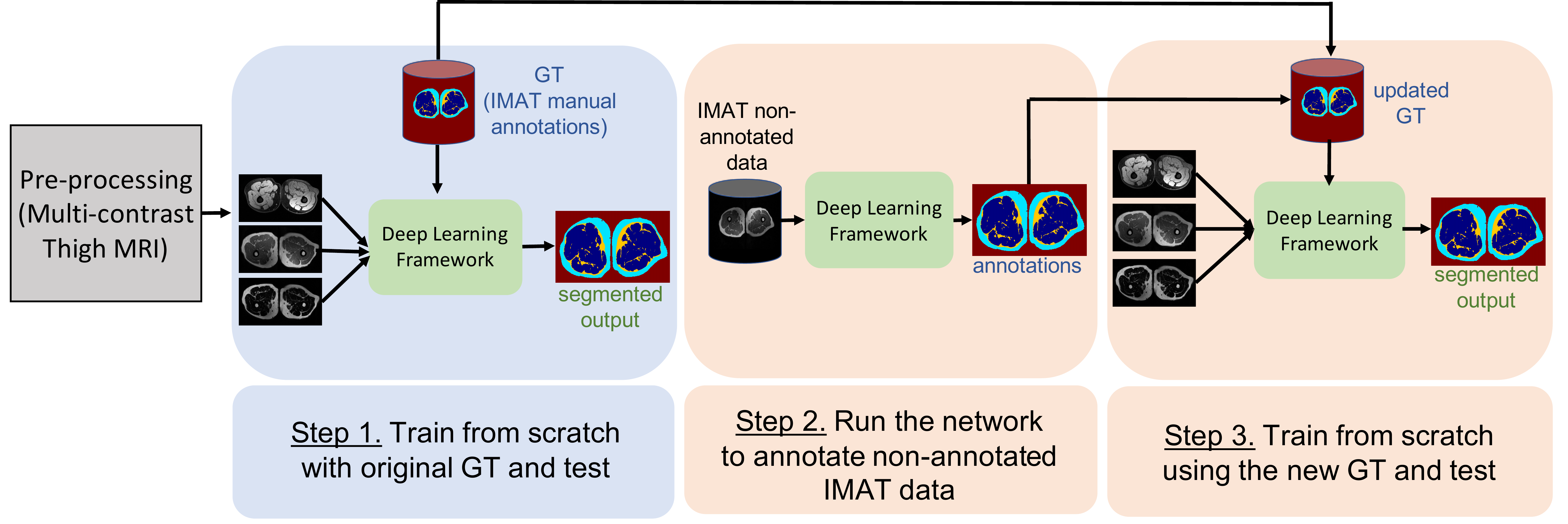}
    \caption{The semi-supervised setting for dealing with scarcely labeled data for IMAT segmentation. GT = ground truth. }
    \label{fig:semi_sup}
\end{figure}

In medical image segmentation, semi-supervised learning is finding wider significance in tackling the scarce availability of labeled data \cite{CHEPLYGINA2019280}. Subcutaneous adipose tissue and IMAT are separated by the anatomical structure called fascia lata. Due to its low-water content, fascia lata has a low signal intensity, and therefore is difficult to visualize or is distorted in MRI scans \cite{doi:10.1002/jmri.25031}. Expert annotation of the fascia lata (and as a result, IMAT identification) is challenging. For example, only very small portion of our dataset have the IMAT annotations by expert radiologists. To take into account the lack of ground truths for IMAT while targeting accurate segmentation of IMAT, we designed a semi-supervised approach (Figure~\ref{fig:semi_sup}). We used the self-training process where the labels were propagated from the labeled data to the unlabeled data. We used a three step process where the unlabeled data was annotated using the proposed network. In the first step, the network was trained with the scarcely labeled data (original ground truth) from experts. The trained model was used to generate new annotations for IMAT (step 2). This new annotated data was added to the originally available ground truth for training and testing (step 3) the system performance. The semi-supervised approach was used to generate segmentation maps for fat, muscle, bone, bone marrow, and IMAT.   \\ 

\subsection{\textbf{Dropouts}} The model capacity, hence complexity, increases quickly when neural networks become deeper and can easily cause over-fitting problems. In this regard, \textit{dropout} was introduced as a regularization technique to sparsify deep networks to avoid over-fitting problems \cite{srivastava2014dropout}. In regular dropout, one of the dominant methods due to its simplicity, neural units are randomly dropped to prevent the feature co-adaptation. The dropout function can be defined as,
\begin{equation}
Y_{l} = (X_{l} \odot \epsilon)W_{l}   \in \epsilon_{i,j} \sim p(\epsilon_{i,j}), 
\end{equation}
where $Y_{l}$ is $L \times M$ output matrix at layer $l$, $X_{l}$ is $L \times K$ input, and $W_{l}$ is the $K \times M$ weight matrix. $\epsilon$ is the noise sampled from a known probability distribution (usually Bernoulli). 
There is a lack of clarity regarding whether prior information can play a significant role in weight selection for the dropout mechanism as opposed to random selection, especially in semi-supervised learning where the lack of large amounts of training data require the selection procedure to be reconsidered carefully. 

One of the major drawbacks of regular dropout is a slower training time~\cite{srivastava2014dropout}. At the same time, the set of dropped weights do not take into account any prior information. It has been shown that the noise can be sampled from a continuous distribution, such as Gaussian, which gives better performance \cite{neelakantan2015adding}. 
Our objective related to dropout design was to increase robustness and speed of the training procedure for segmentation, and provide an improved segmentation performance through better generalization of the network. To achieve this, we have adapted variational and targeted dropouts into the Tiramisu network to obtain benefits in convergence and robustness.\\ 


\noindent\textbf{Variational Dropout (VD):} As proposed in \cite{Kingma:2015:VDL:2969442.2969527}, VD is a generalization of Gaussian dropout where the dropout rates are learned. In our semi-supervised learning paradigm, since we have limited data, we argue that learning model parameters (such as dropout rates) can lead to better models. In other words, instead of randomly picking weights to be dropped in a regular network, we propose to use VD to achieve better efficiency and robustness.
More specifically, given the input and output datasets $X=\{x_1,x_2,..,x_N\}$ and $Y=\{y_1,y_2,..,y_N\}$, we seek the posterior distribution $p(\omega|X,Y)$, by which we can predict output $y^*$ for a new input point $x^*$ by solving the integral 
\begin{equation}
    p(y^*|x^*,X,Y) = \int p(y^*|X^*,\omega)p(\omega|X,Y)d\omega.
\end{equation}
To obtain the posterior distributions, a Gaussian prior distribution $N(0, I)$ can be placed over the network weights leading to much faster convergence. The objective function is given as,
\begin{equation}
\max_{w,\alpha} L_{(w,\alpha)} -D_{KL}(q(W) || p(W)),
\end{equation}
where $D_{KL}$ is the Kullback Leibler divergence loss, p(W) is the prior distribution (a Gaussian), q(W) is the posterior distribution, and $\alpha$ is the variance of the Gaussian distribution.\\ 

\noindent\textbf{Targeted Dropout (TD):} A post-hoc pruning mechanism was proposed in \cite{gomez2019learning}, with the rationale to make pruning part of the learning process for sparsity regularization for classification problems. Given a neural network parameterized by $\Theta$, the goal was to find the optimal parameters $W_{\Theta}$ such that the loss ($L(W_{\Theta})$) was minimized. For efficiency and generalization reasons, $|W_{\Theta}| \leq k $, where only $k$ weights of the highest magnitude in the network were employed. In this regard, a deterministic approach is to drop the lowest $|W_{\Theta}|-k$ weights. In TD, using a target rate $\gamma$ and a dropout rate $p$, first a target set $\textit{T}$ was generated with the lowest weights with the target rate $\gamma$. Next, weights were stochastically dropped from the target set $\textit{T}$ with the dropout rate $p$. In our study, we applied this strategy to pixel level classification where five different tissues were segmented in thigh MRI.  

\section{Results}\label{ER}


\subsection{\noindent\textbf{Deep Network Parameters}} 

The Tiramisu network was used as a baseline segmentor, which contains $103$ layers, composed of approximately $9$ million trainable parameters, and $35$ $million$ neurons. We used a soft-max cross-entropy loss function and the learning rate was set to $0.00005$ for initialization. Although the \textit{adam} optimizer was used with Xavier initializer, the bias term was initialized to zero. The ReLU activation function was used along with batch normalization, and the batch size was set to $3$. For the dense blocks, we empirically found that a growth rate of $24$ was comparably better than others. The experiments were conducted on $2$ Nvidia Titan-XP GPUs each with $12GB$ memory. We evaluated the experiments using a $5$-fold cross-validation strategy. For the supervised method, the split ratio of the train, validation, and test sets was set to $70:10:20$. For the semi-supervised method, in the first step, the train and validation split ratio was set to $90:10$, while in the second step (similar to the supervised approach) the train, validation, and test split ratio was set to $70:10:20$.

\subsection{\textbf{Supervised-Learning}} The segmentation results using supervised learning were evaluated with conventional dice score, Hausdorff distance, sensitivity, and specificity metrics. The results are presented in Table \ref{tab:my_label}. We analyzed the performance for $2$-, $4$-, and $5$- tissue segmentation and found that all of our proposed methods performed statistically similarly ($p>0.005$) when the system is fully supervised. For 5-tissue segmentation (including IMAT), a semi-supervised approach was used. With a greater number of tissues in the segmentation task, the performance of muscle segmentation decreases. This may be explained by the increased number of the classes to learn and also the increased correlation of the tissues classified. In contrast, fat tissue does not have direct spatial correlation with bone and bone marrow tissues. The results were also compared with a baseline method (U-net) using multi-contrast MRIs. The method preforms well in some of the performance parameters, but our proposed supervised methods outperforms in majority of the cases in dice scores (fat and bone-marrow), sensitivity (muscle, fat, and bone-marrow), and specificity (muscle, fat, bone, and bone-marrow). Whereas, our proposed semi-supervised method outperforms U-net in all performance parameters with a significant margin for all tissues except for bone sensitivity which is slightly lower. This highlights the fact that it was difficult to segment all tissues consistently with high performance, which we have achieved here with our proposed semi-supervised method.   

Results for regular dropout are reported for 2-tissue segmentation, whereas for 4-tissue segmentation all three dropout results (regular (R), variational dropout (VD), and targeted dropout (TD)) are reported. We observed that results for various parameters were mixed when choosing among the dropouts. In general, Tiramisu with variational dropout (Tiramisu+VD) performance was either better or comparable to Tiramisu with regular (Tiramisu+R) and targeted dropout (Tiramisu+TD). A significant benefit of these dropouts was also observed in the convergence rate and for semi-supervised learning. We concluded that variational dropout gives better performance than regular and targeted dropout in thigh tissue segmentation and used it in the semi-supervised setting.    

\begin{table}[!h]
    \centering
    \caption{Summary of the segmentation performance for 4- and 5-tissue segmentation of thigh MRIs (DSC = dice score).}
    \resizebox{!}{0.25\textwidth}{
    \begin{tabular}{c|c|c|c|c|c|c}
        \textbf{Method} & \textbf{Muscle} & \textbf{Fat} & \textbf{Bone} & \textbf{Bone Marrow} & \textbf{IMAT}  &\textbf{Metric (\%)} \\
        \hline
        
        \multirow{3}{*}{\textbf{\thead{5-Tissue Segmentation\\(Proposed Semi-supervised)}}} & $97.52$ &  $94.61$ &  $95.93 $ & $96.83$ & $80.14 $ & DSC \\ \cline{2-6}
        & $97.11$ & $92.89$ & $99.36$ & $95.69$ &$88.15$ &  Sensitivity \\ \cline{2-6}
        & $99.73$ & $99.79$ & $99.96$ & $99.99$ & $99.44$ & Specificity \\ 
        \midrule
        \multirow{3}{*}{\textbf{\thead{2-Tissue Segmentation\\(Tiramisu+R)}}} & $94.67$ & $90.92$ &  -   &- &-  & DSC \\ \cline{2-6}
        &99.75 & $91.29$&  - &-  &- & {Sensitivity} \\ \cline{2-6}
        &99.71&  $93.64$ & - & - &  -& {Specificity} \\
        \midrule
        \multirow{3}{*}{\textbf{\thead{4-Tissue Segmentation\\(U-net multi-contrast)}}} & $87.37$ &$89.89$ &  $89.26$ & $86.01$ &- & DSC \\ \cline{2-6}
        & $86.99$ & $93.20$ & $88.81$ & $88.70$ & - & Sensitivity \\ \cline{2-6}
        & $98.99$ & $98.61$ & $99.98$ & $99.94$ & -& Specificity \\
        
        \midrule
        
        \multirow{3}{*}{\textbf{\thead{4-Tissue Segmentation\\(Tiramisu+R)}}} & $86.65$ &$92.96$ &  $80.84$ & $88.49$ &- & DSC \\ \cline{2-6}
        & $86.99$ & $94.14$ & $78.90$ & $89.53$ & -& Sensitivity \\ \cline{2-6}
        & $98.54$ & $99.16$ & $99.99$ & $99.96$ & -& Specificity \\
        
        \midrule
        
        
        \multirow{3}{*}{\textbf{\thead{4-Tissue Segmentation\\(Tiramisu+VD)}}} &
        $86.85$ & $93.01$ & $83.62$ & $84.30$ & -& DSC \\ \cline{2-6}
        & $90.30$ & $93.20$ & $85.08$ & $83.42$ & -& Sensitivity \\ \cline{2-6}
        & $98.26$ & $99.25$ & $99.98$ & $99.97$ &- & Specificity \\
        \midrule
        
        
        \multirow{3}{*}{\textbf{\thead{4-Tissue Segmentation\\(Tiramisu+TD)}}} & $80.48$ & $92.03$ & $82.01$ & $79.97$ &- & DSC \\ \cline{2-6}
        & $73.78$ & $95.70$ & $83.36$ & $72.72$ & -& Sensitivity \\ \cline{2-6}
        & $99.09$ & $98.71$ & $99.98$ & $99.98$ & -& Specificity \\
        \midrule
        
        \bottomrule
    \end{tabular}}
    \label{tab:my_label}
\end{table}


\begin{figure}[!t]
\centering
\begin{tabular}{c|ccc}
\hline
& (A) & (B) & (C) \\ \hline  

\textbf{\thead{Water and \\Fat}} &\includegraphics[height =2cm,width=3.5 cm]{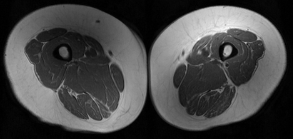} & \includegraphics[height =2cm,width=3.5 cm]{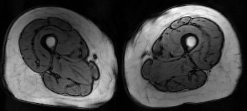}
& \includegraphics[height =2cm,width=3.5 cm]{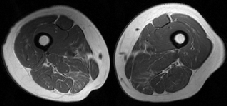} \\

\textbf{\thead{Fat \\Suppressed}} &\includegraphics[height =2cm,width=3.5 cm]{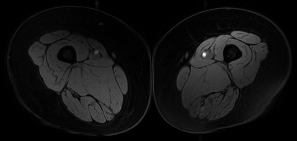} & \includegraphics[height =2cm,width=3.5 cm]{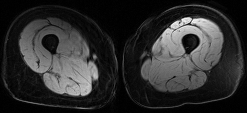} &
\includegraphics[height =2cm,width=3.5 cm]{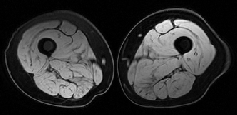} \\

\textbf{\thead{Water \\Suppressed}} & \includegraphics[height =2cm,width=3.5 cm]{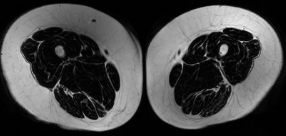} & \includegraphics[height =2cm,width=3.5 cm]{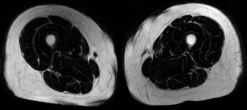} &
\includegraphics[height =2cm,width=3.5 cm]{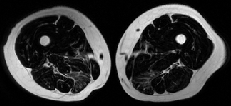} \\

\textbf{\thead{GT}}& \includegraphics[height =2cm,width=3.5 cm]{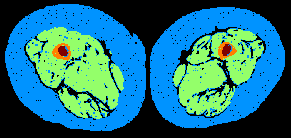} & \includegraphics[height =2cm,width=3.5 cm]{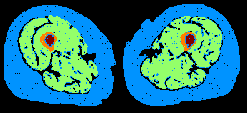} & \includegraphics[height =2cm,width=3.5 cm]{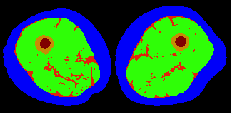} \\

\textbf{\thead{Prediction}}&\includegraphics[height =2cm,width=3.5 cm]{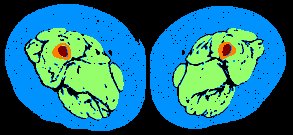} & \includegraphics[height =2cm,width=3.5 cm]{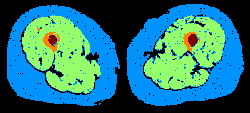} & \includegraphics[height =2cm,width=3.5 cm]{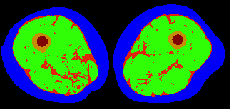}

\\ \hline
\end{tabular}
\caption{Multi-contrast thigh MRI scans with GT and predictions for tissue segmentation. \underline{(A)} Muscle, fat, bone, and bone marrow segmentation using U-net. \underline{(B)} Muscle, fat, bone, and bone marrow segmentation using our proposed method. \underline{(C)} Muscle, fat, bone, bone marrow, and IMAT segmentation using our proposed semi-supervised method.} 
\label{fig:result_scans}
\end{figure}

\subsection{\noindent\textbf{Semi-supervised Learning}}
Experiments were conducted for 5-tissue segmentation including fat, muscle, bone, bone marrow, and IMAT using variational dropout. For IMAT, annotating the fascia lata is a labor-intensive task. For semi-supervised learning, images from $20$ subjects were manually segmented using the Amira software (AMIRA, Mercury Computer Systems, Berlin, Germany) by participating experts. We first trained the Tiramisu network with the expert annotated scans for semi-supervised learning. Next, we used the trained network to automatically label the rest of the non-annotated scans (from 20 subjects), which were then combined with the manually annotated ground truths for re-training from scratch. The whole data set was then divided into $80\%$ and $20\%$ portions for training (images from 40 subjects) and testing (images from 20 subjects), respectively. The results for performance parameters using the semi-supervised approach are given in Table \ref{tab:my_label}. We obtained mean dice scores of $97.52 \%$, $94.61\%$, $95.63\%$, $96.83 \%$, and $80.14\%$ for muscle, fat, bone, bone marrow, and IMAT tissues, respectively. The sensitivity value for the IMAT tissue was lower than that for other tissues, which revealed that IMAT remains the most challenging tissue to be segmented within the thigh region. We also achieved significantly higher dice scores for muscle, fat, bone, and bone marrow tissues using the semi-supervised approach, even higher than the 2- and 4-tissue segmentation using the fully supervised setting. This demonstrates the effectiveness of our proposed semi-supervised setting for these tissues (muscle, fat, bone, and bone marrow) as well as the challenging IMAT tissue. The segmented images using our proposed methodology are shown in Figure~\ref{fig:result_scans} for both fully supervised U-net (first row) method, the Tiramisu (second row) method, and semi-supervised (third row) technique, along with the ground truth (GT) segmented images and the corresponding input of multi-contrast MR scans. The input MRIs include fat suppressed (MRI1), water and fat (MRI2), and water suppressed (MRI3) contrasts. The predictions include 4-tissues (muscle, fat, bone, and bone marrow) in the first and second rows, and 5-tissues (muscle, fat, bone, bone marrow, and IMAT) in the third row.

\newgeometry{margin=1cm}

\begin{sidewaystable*}
    \centering
    
    \begin{tabular}{|c|c|c|c|c|c|c|c|c|c|c|c|c|c|}
        \hline
        \multirow{2}{*}{\textbf{Method}} & \multirow{2}{*}{\textbf{Input}} & \multicolumn{3}{|c|}{\textbf{Muscle}} &\multicolumn{3}{|c|}{\textbf{Fat}} &   \multicolumn{3}{|c|}{\textbf{Bone}} & \multicolumn{3}{|c|}{\textbf{Bone Marrow}}  \\
        \cline{3-14}

        & & DSC & Sensitivity & Specificity & DSC & Sensitivity & Specificity& DSC & Sensitivity & Specificity& DSC & Sensitivity & Specificity \\
        \hline
        
        \toprule
        
        \multirow{4}{*}{U-net}& MRI1 & 75.96 & 72.94 & 97.93 & 82.26 & 87.92 & 95.37 & 71.74 & 70.04 & 99.97& 66.79 & 67.03 & 99.87 \\
        \cline{2-14}
        
        & MRI2 & 83.64 & 81.21 & 98.98 & 90.82 & 91.61 & 99.08& 91.04 & 91.70 & 99.98& 88.20 & 89.92 & 99.95 \\
        \cline{2-14}
        
        & MRI3 & 86.01 & 85.51 & 98.88 & 91.56 & 89.64 & 99.39& 81.55 & 83.77 & 99.96& 76.69 & 78.99 & 99.91 \\
        \cline{2-14}
        
        & Multi-contrast & 87.39 & 86.99 & 98.99 & 89.89 & 93.20 & 98.61& 89.26 & 88.81 & 99.98& 86.06 & 88.70 & 99.94 \\
        \hline
        \midrule
        
        \multirow{4}{*}{Proposed}& MRI1 & 80.19 & 73.30 & 98.61 & 85.77 &83.25 & 98.64 & 75.44 & 79.20 & 99.95& 74.62 & 79.27 & 99.88 \\
        \cline{2-14}
        
        & MRI2 & 78.71 & 83.54 & 97.06 & 84.74 & 81.33 & 98.97& 91.48 & 90.94 & 99.98& 90.10 & 93.57 & 99.95 \\
        \cline{2-14}
        
        & MRI3 & 84.98 & 89.63 & 97.78 & 91.59 & 88.82 & 99.54& 74.84 & 74.96 & 99.96& 72.39 & 75.05 & 99.89 \\
        \cline{2-14}
        
        & Multi-contrast & 86.85 & 90.30 & 98.26 & 93.01 & 93.20 & 99.25& 83.62 & 85.08 & 99.98& 84.30 & 83.42 & 99.97 \\
        \cline{2-14}
        
        
        \bottomrule
    \end{tabular}
    \caption{Summary of the segmentation performance using single and multi-contrast MRIs.}
    \label{tab:my_label_semi}
\end{sidewaystable*}

\restoregeometry
 
\section{Discussion} \label{dcon}
The segmentation of multiple tissues in thigh MRI scans is a challenging task due to lack of annotated data and the complex nature of the tissue distribution in the thigh region. In this study, we have presented a semi-supervised deep learning to train segmentation network (Tiramisu) for overcoming the limited availability of annotated data while segmenting multiple tissues with high accuracy. Tiramisu architecture included 103 dense layers; hence, it provided a comprehensive combination of low and high level imaging features for pixel level labeling with high precision. 
In our experiments, we have used the U-Net \cite{ronneberger2015u} architecture to get the baseline results. In medical image segmentation, U-Net has been widely used for a variety of radiology images and has been found to perform well for MRI. It derives its name form the U-shaped architecture which follows an encoder-decoder structure and differs from conventional convolutional neural networks (CNNs). We have further experimented to verify the benefits of using multi-contrast MRI. In our experiments, we have segmented thigh tissues using single-contrast as well as multi-contrast MRI scans with both U-Net and the proposed architectures (Tiramisu) with results presented in Table \ref{tab:my_label_semi}.        


\begin{figure}[!t]
\centering
\begin{tabular}{c|c|c}

Image & U-net & Tiramisu \\ \hline \hline
\textbf{GT} & \includegraphics[width=3 cm]{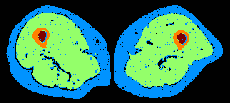} & \includegraphics[width=3 cm]{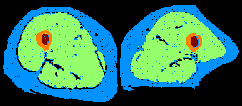} \\  \hline
\textbf{MRI1} & \includegraphics[width=3 cm]{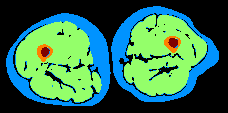} & 
\includegraphics[width=3 cm, height = 1.48cm]{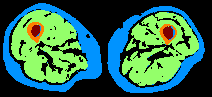} \\ \hline
\textbf{MRI2} & \includegraphics[width=3 cm]{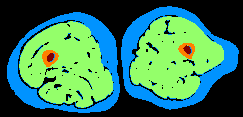} &
\includegraphics[width=3 cm, height = 1.48cm]{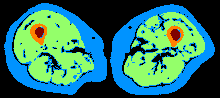} \\ \hline
\textbf{MRI3} & \includegraphics[width=3 cm]{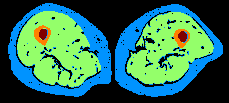} &
\includegraphics[width=3 cm]{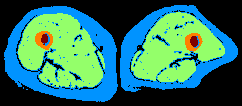}\\ \hline
\textbf{multi-contrast} & \includegraphics[width=3 cm]{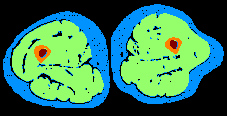} &
\includegraphics[width=3 cm, height = 1.48cm]{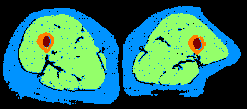} \\ 
\hline
 

\end{tabular}
    \caption{Examples of ground truth and segmented images with MRI1 (fat-suppressed), MRI2 (water-and-fat), MRI3 (water-suppressed) and multi-contrast inputs using \underline{first column} U-net and \underline{second column} Tiramisu architectures.}
    \label{fig:single_contrast_MR}
\end{figure}

\begin{figure}[!ht]
\centering
\begin{tabular}{cc}

\includegraphics[width=56mm]{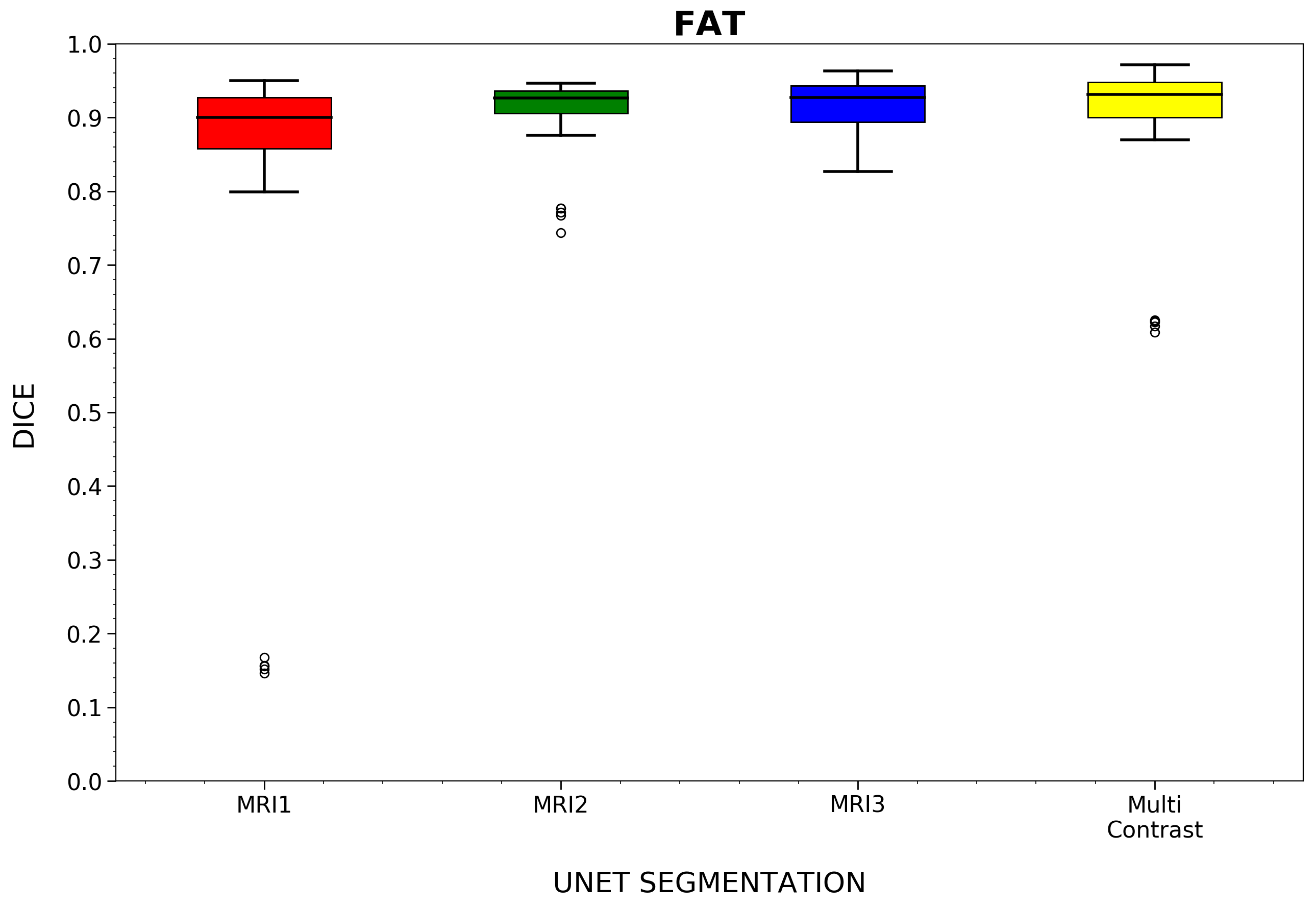} & \includegraphics[width=56mm]{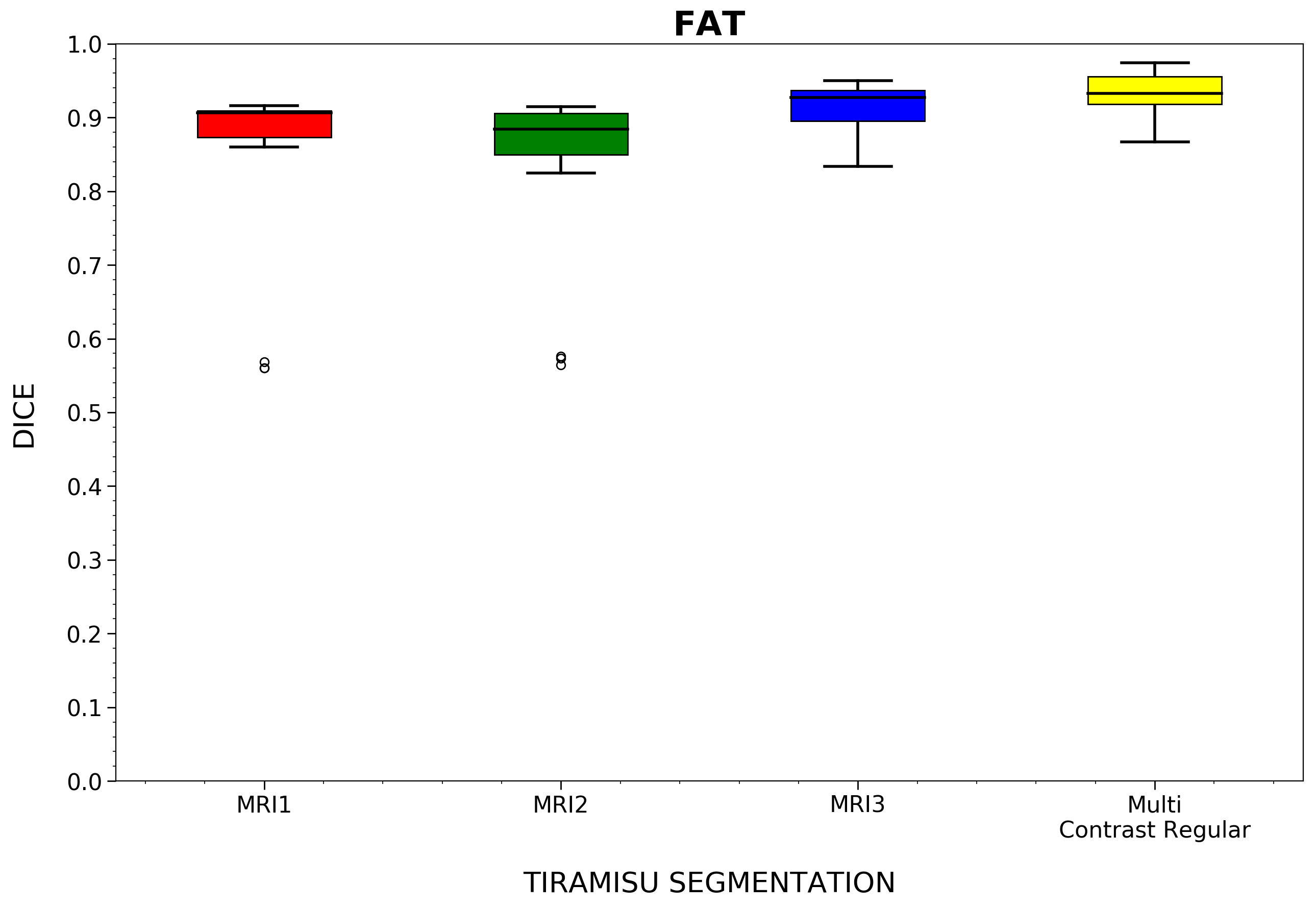} 
 \\ \hline
 
  \includegraphics[width=56mm]{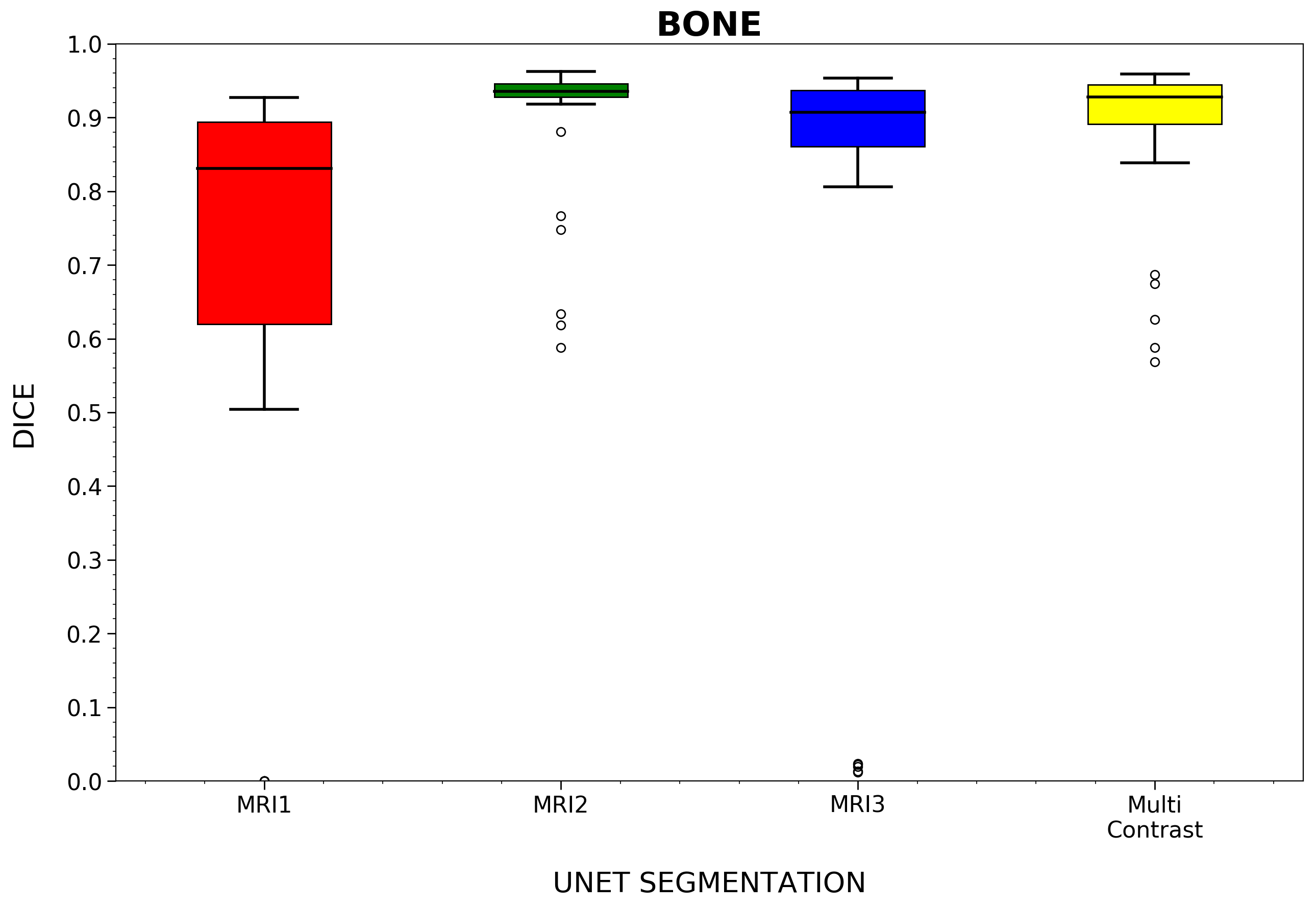} & \includegraphics[width=56mm ]{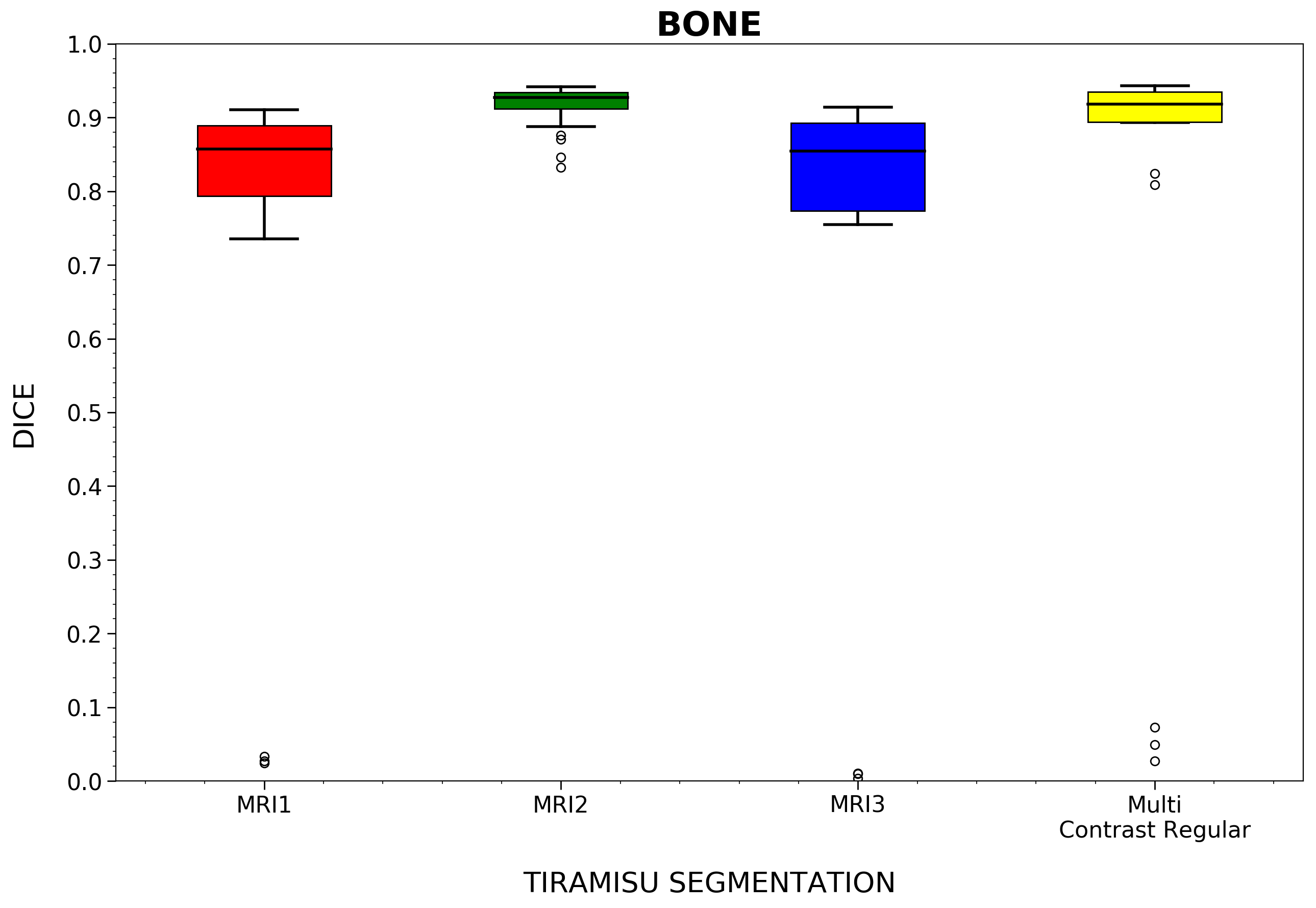} 
  \\ \hline
 
 \includegraphics[width=56mm]{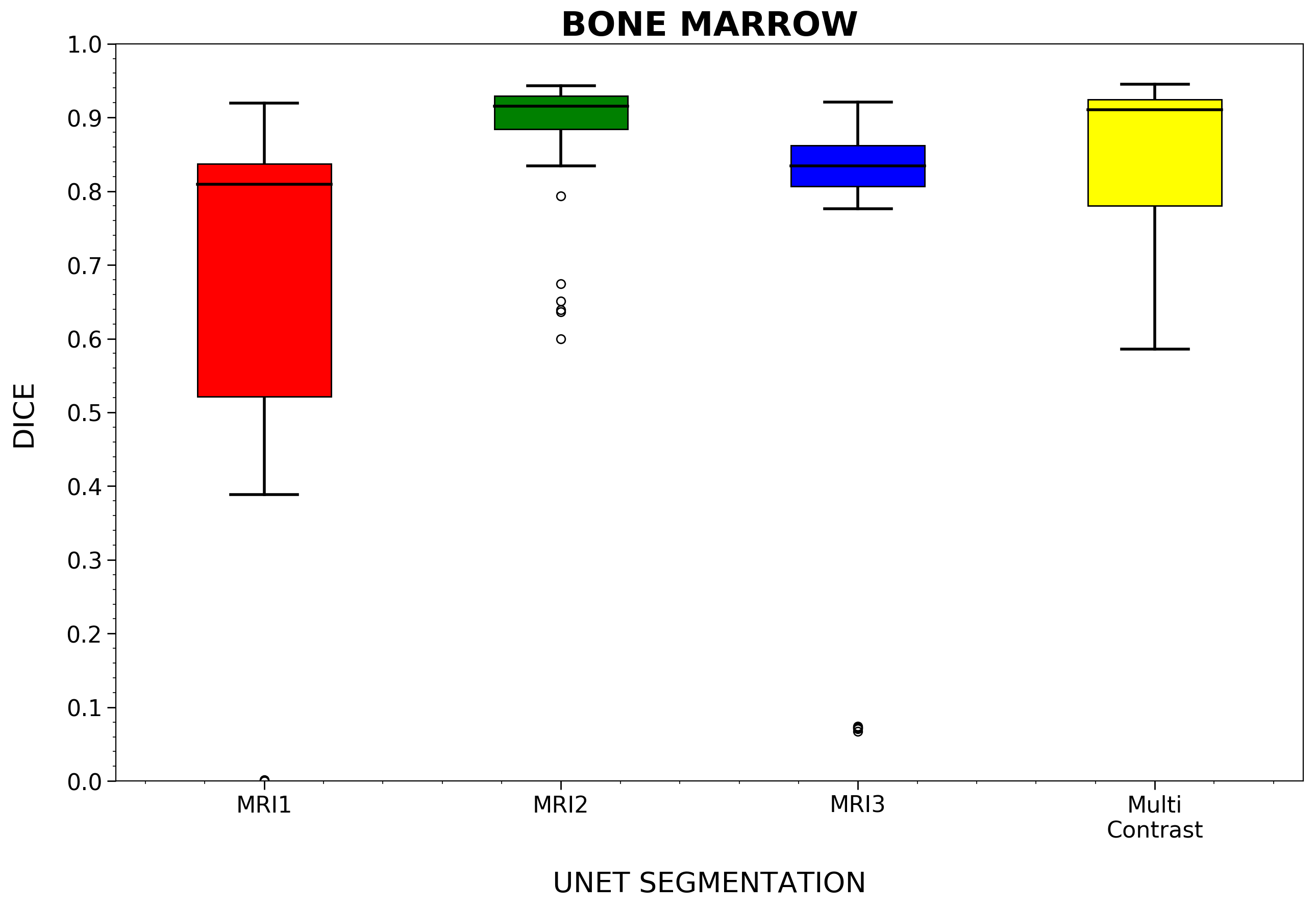} & \includegraphics[width=56mm ]{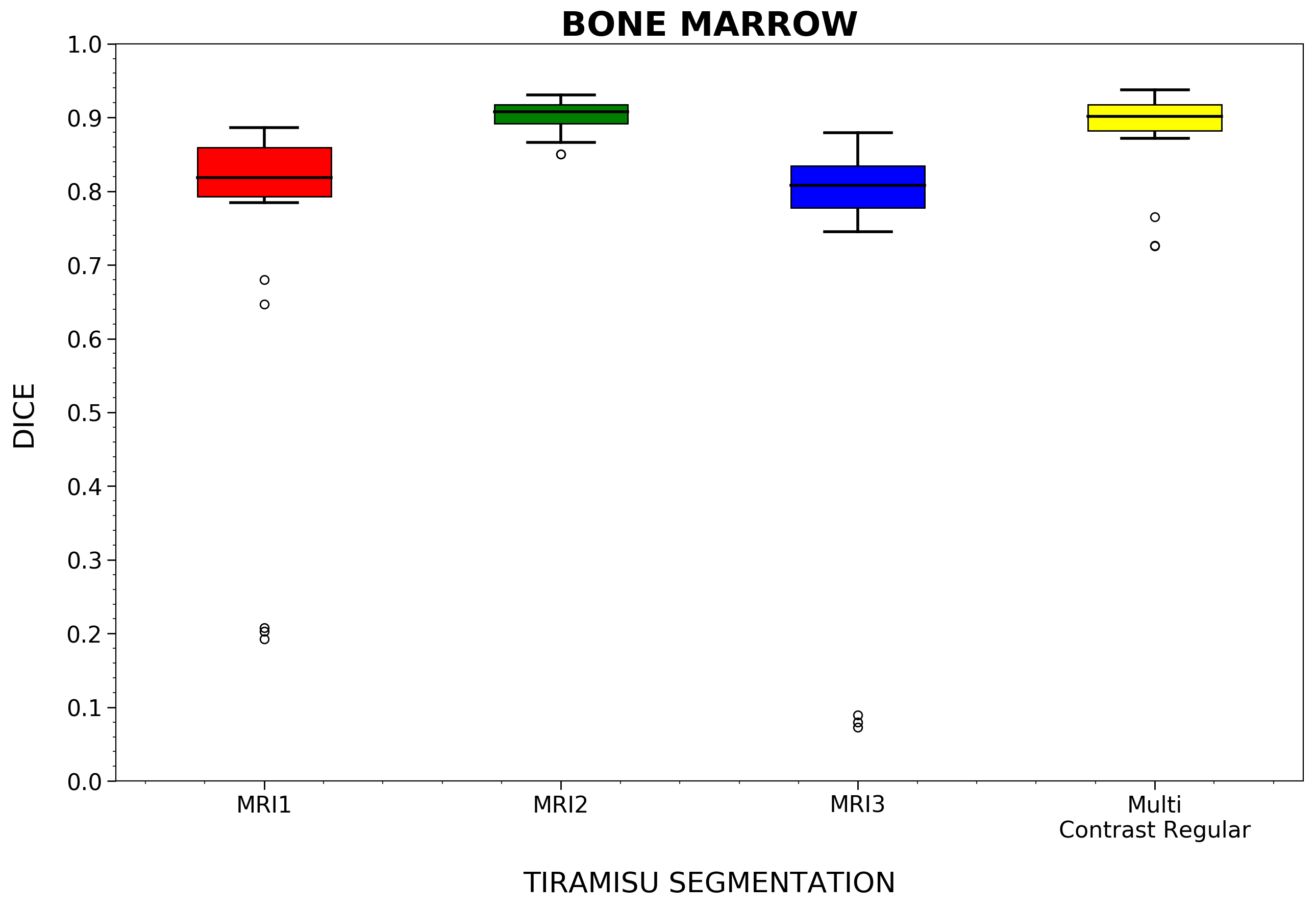} 
 \\ \hline
 
 \includegraphics[width=56mm]{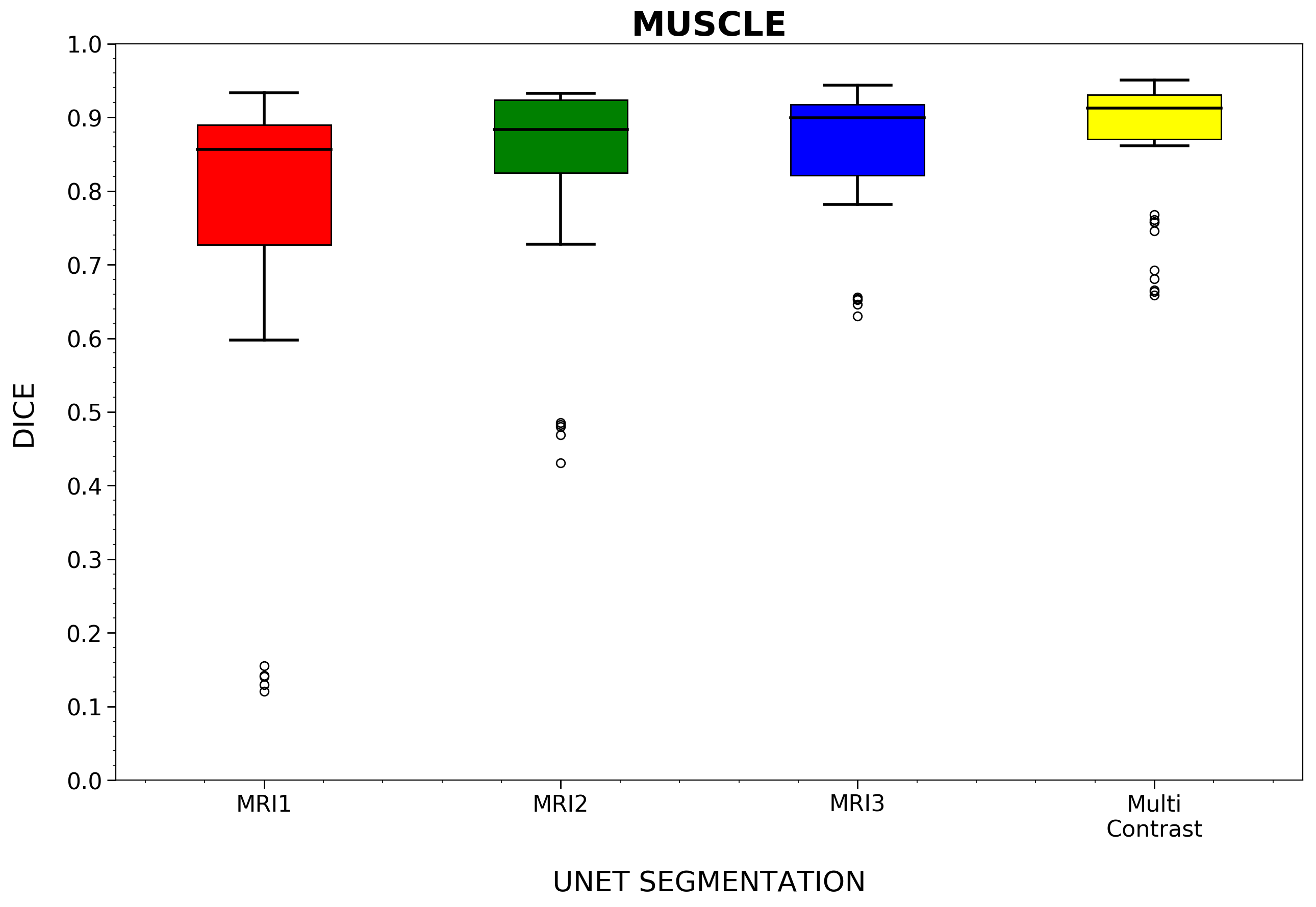} & \includegraphics[width=56mm ]{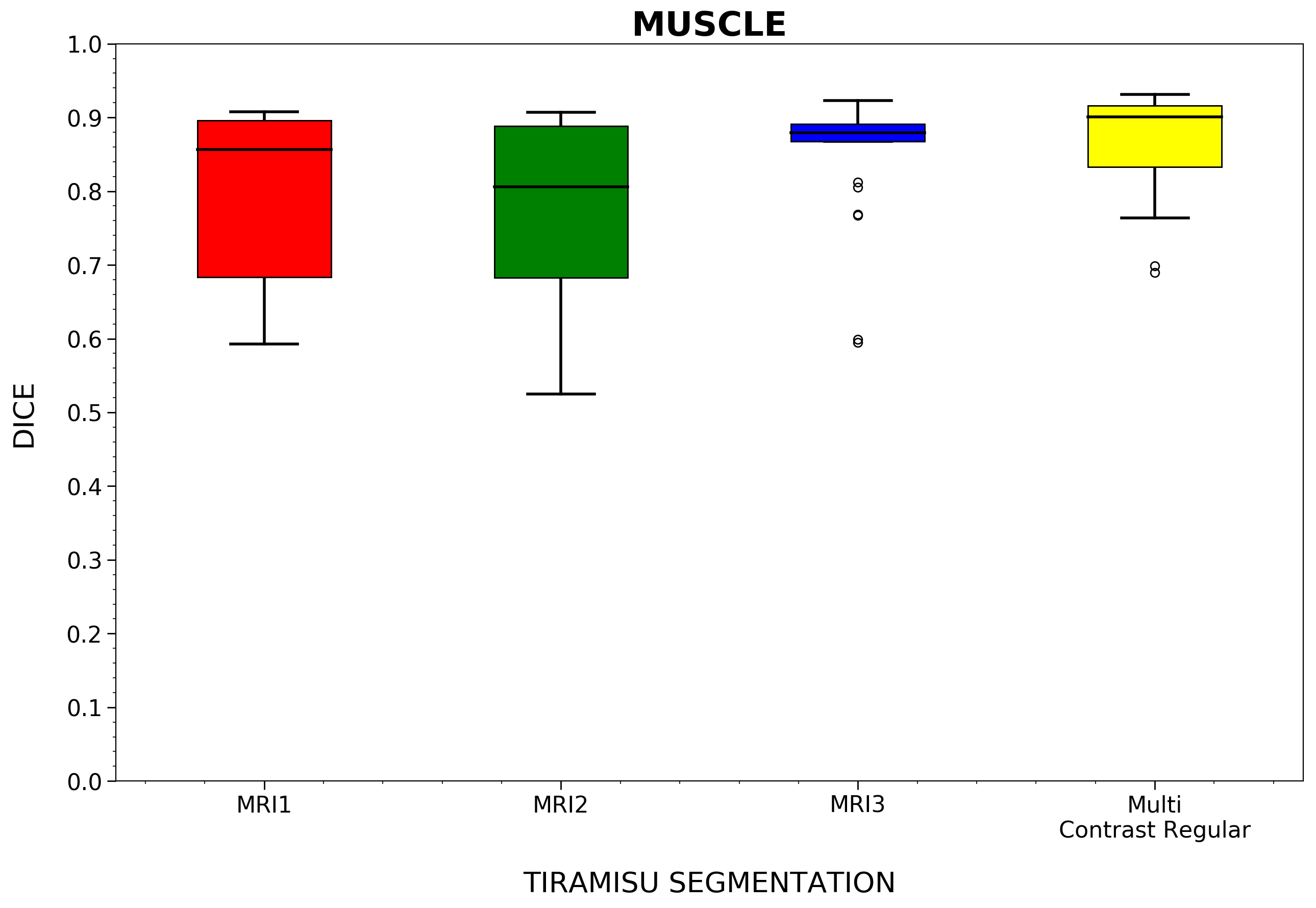} 
 
\end{tabular}
\caption{Box plots showing a comparison of dice scores for muscle, fat, bone, and bone marrow tissues using: \underline{First column} U-net with fat-suppressed (MRI1), water-and-fat (MRI2), water-suppressed (MRI3), and multi-contrast inputs. \underline{Second column} Tiramisu with fat-suppressed (MRI1), water-and-fat (MRI2), water-suppressed (MRI3), and multi-contrast for inputs.}
    \label{fig:performance_parameters}
\end{figure}

\begin{figure}[!t]
\centering
\begin{tabular}{cc}

 \includegraphics[width = 75mm]{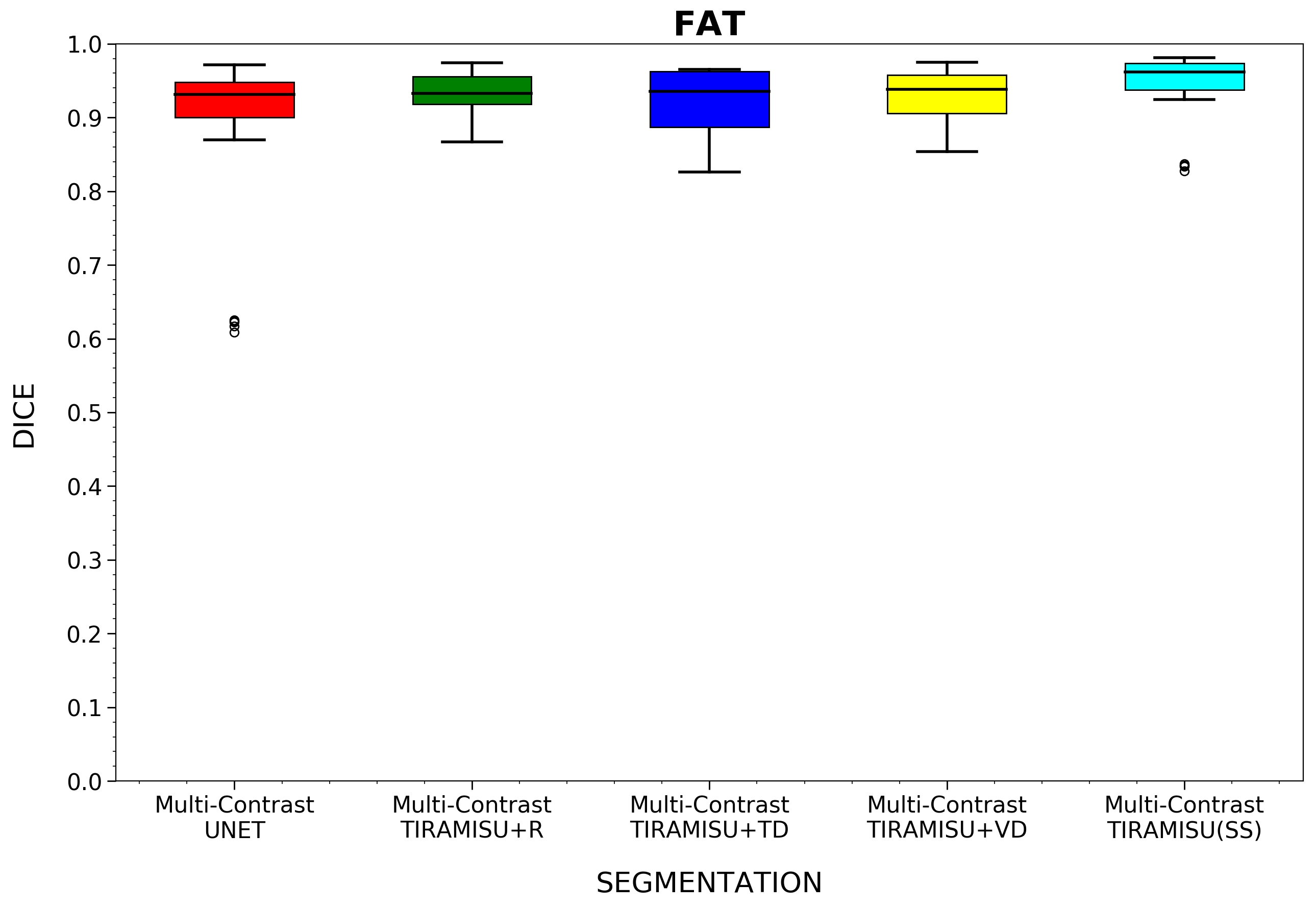} 
 
 & \includegraphics[width = 75mm]{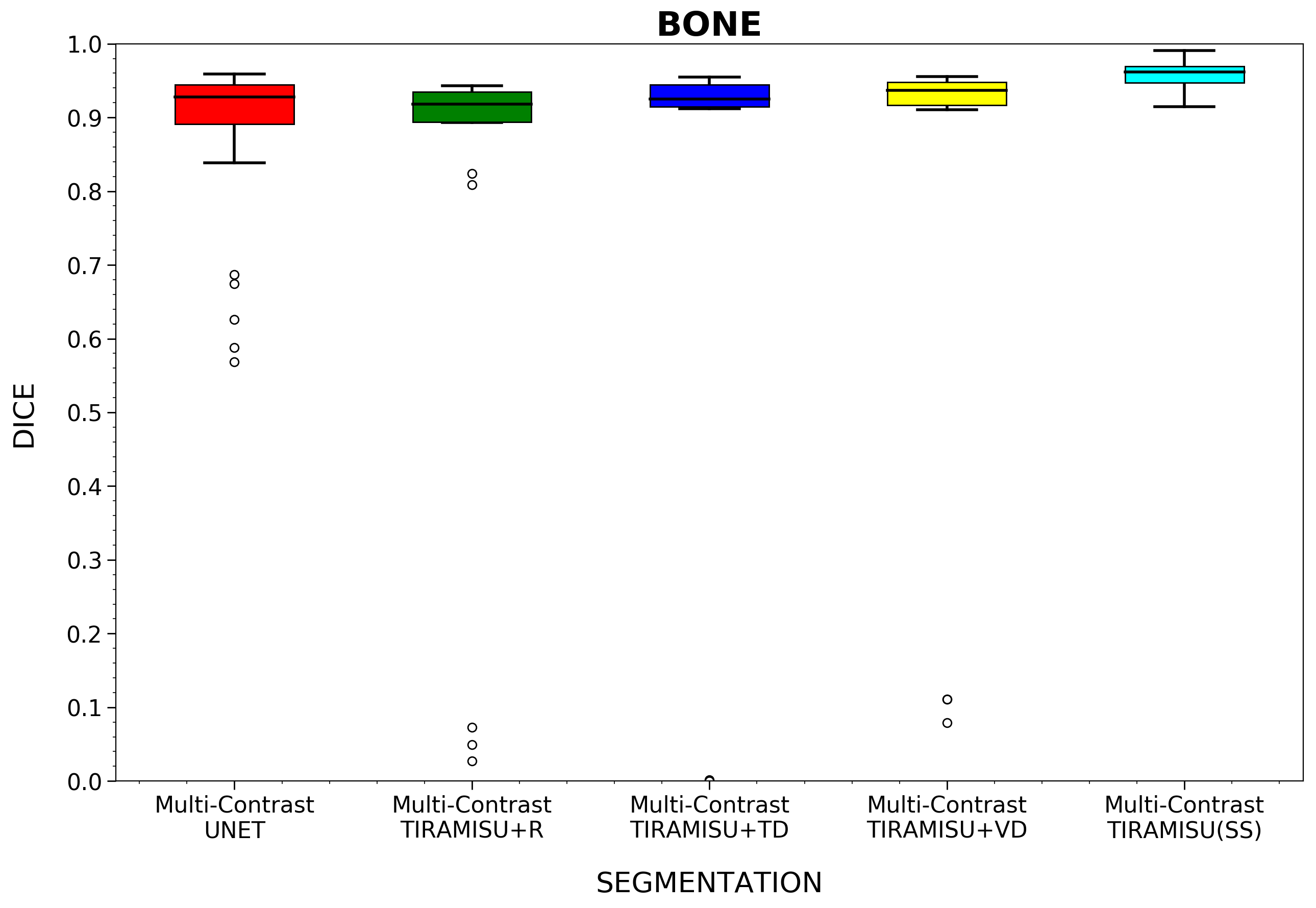} \\ \hline
 
  \includegraphics[width = 75mm]{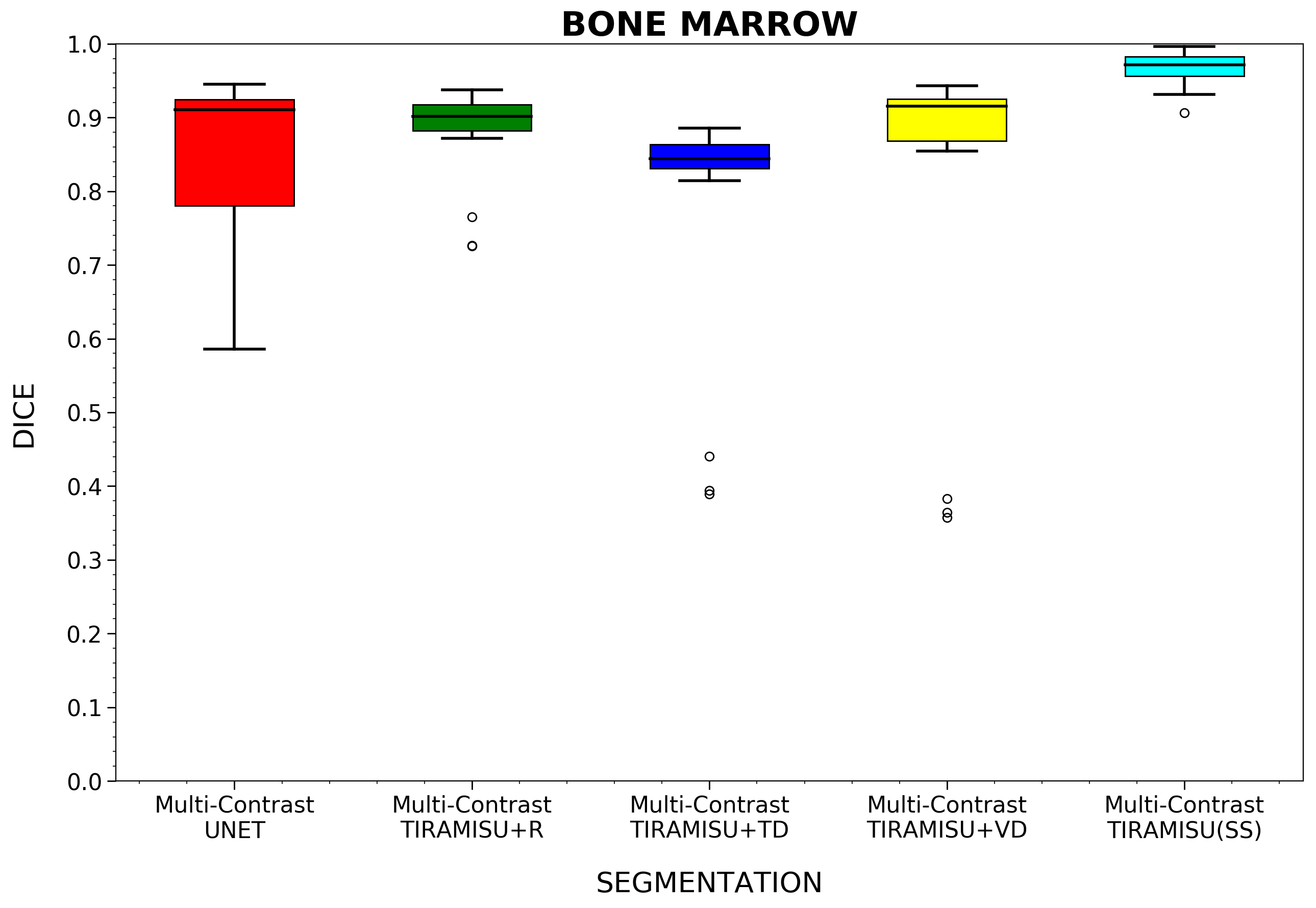} 
  
  & \includegraphics[width = 75mm]{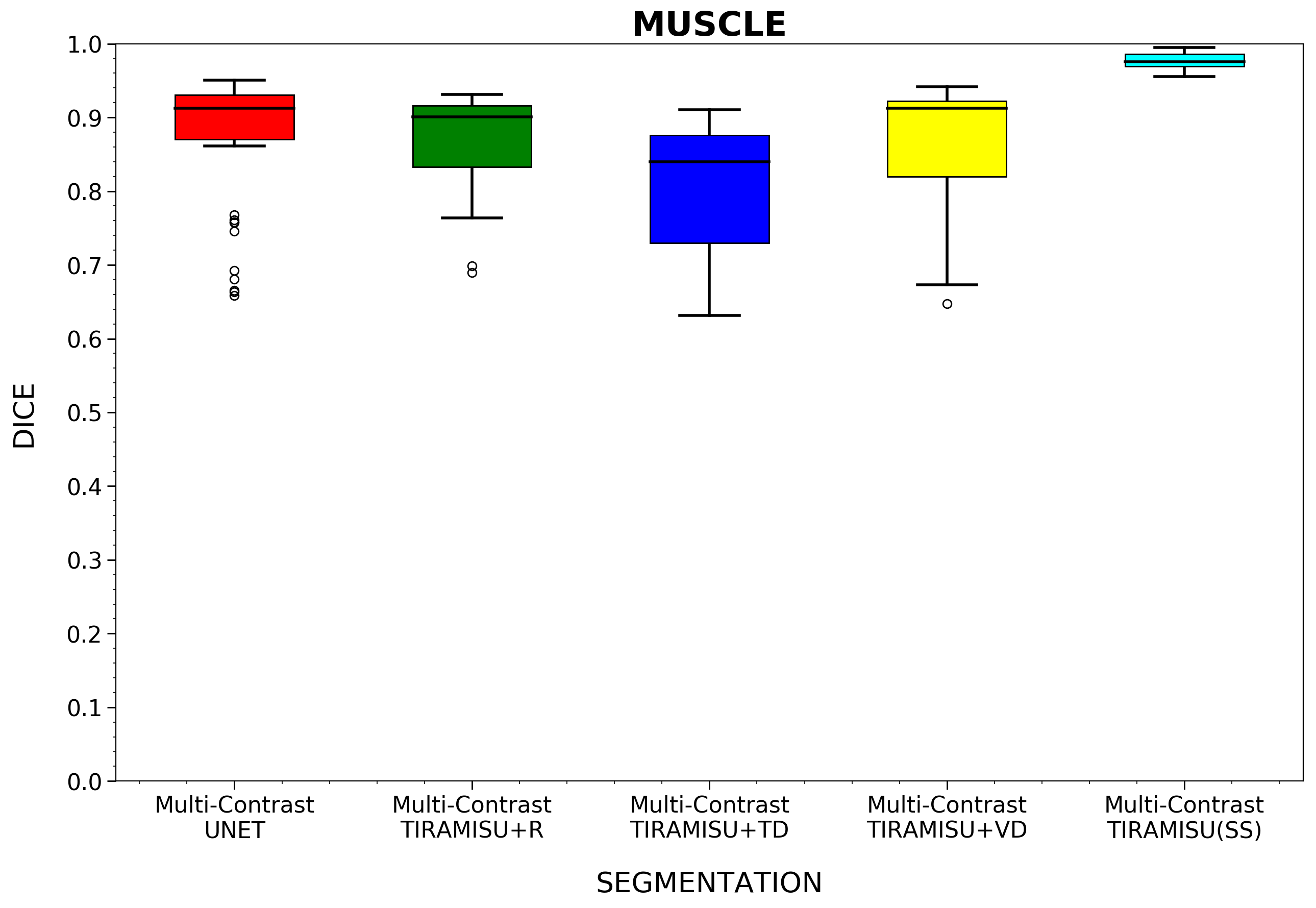} 
 
\end{tabular}
\caption{Box plots showing a comparison of dice scores for muscle, fat, bone, and bone marrow tissues using multi-contrast inputs to U-net and different variants of our proposed methods.}
\label{fig:performance_parameters_all}
\end{figure}

\begin{figure}[!ht]
    \centering
    \includegraphics[width =100mm]{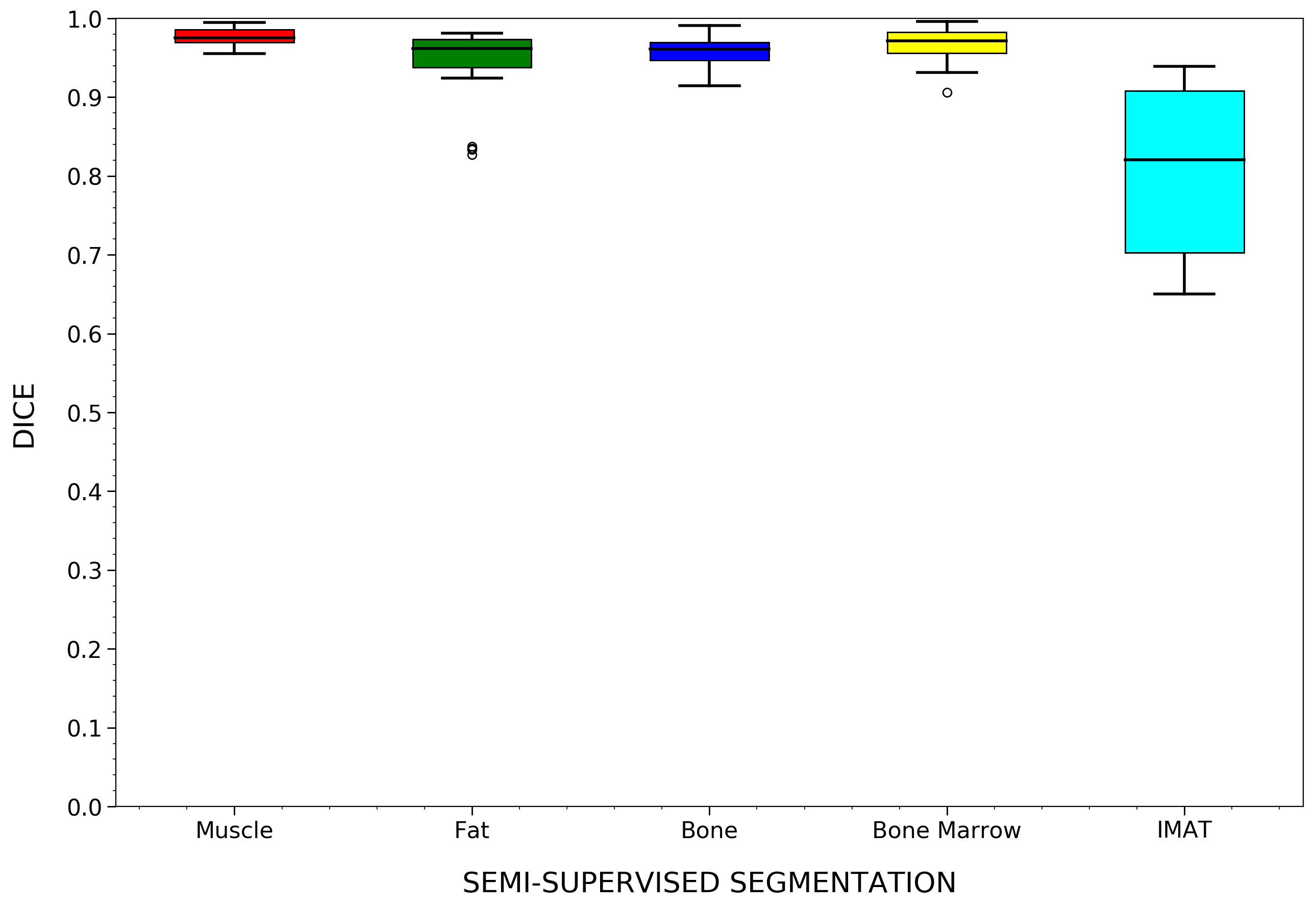}
    \caption{Box plots showing dice scores for various tissues segmented using our proposed semi-supervised method.}
    \label{fig:ssdice}
\end{figure}

\subsection{Multi-contrast MRI Performs Better in Quantification}
We have presented the segmentation of tissues in the thigh using multi-contrast MRI for the first time to the best of our knowledge. We compared the results for segmentation using each single contrast as well (i.e., using fat- suppressed, water-suppressed, and water and fat signal intensity MRIs individually). The results were evaluated using U-net (as a baseline) and the proposed architecture. A summary of the performance parameters are given in Table \ref{tab:my_label_semi}. It is generally observed that multi-contrast MRIs as input to both U-Net and our proposed Tiramisu architectures perform better in most instances. For bone and bone marrow, using MRI2 (water and fat) as the input gives the highest dice scores, whereas for muscle and fat, the performance is better with multi-contrast MRIs. These results are consistent with what we observed earlier (Table \ref{tab:my_label}) where it was difficult to find a single method that outperforms all other methods in this challenging task. It is important to note that with multi-contrast inputs to our proposed Tiramisu architecture, the performance is consistently high for all tissues. Based on these observations, we selected multi-contrast Tiramisu for our proposed semi-supervised method, which has significantly higher performance than all other methods we have compared here. Moreover, the results for the segmentation were compared with ground-truth labels using Hausdorff distance (HD) at 95\% (in mm) \cite{huttenlocher1993comparing}. An average HD value of 1.10 (mm) was achieved using 95 percentile for the proposed semi-supervised framework. The values for HD at 95 percentile for other methods used are presented in Table \ref{tab:HD}.

\begin{table}[!t]
    \centering
    \begin{tabular}{l|c}
         \toprule
         Method &  HD95 (mm) \\ \hline 
         \midrule
         Multi-contrast with U-NET & 3.33 \\
         Multi-contrast with Tiramisu + R & 4.27\\
         Multi-contrast with Tiramisu + VD & 2.81\\
         Multi-contrast with Tiramisu + TD & 3.46\\
         \textbf{Multi-contrast with semi-supervised} & \textbf{1.10}\\
         \hline
         \bottomrule
    \end{tabular}
    \caption{Hausdorff distance based segmentation evaluations. Lower values indicate better performance.}
    \label{tab:HD}
\end{table}
A few examples of segmented images (4-tissue) using multi-contrast and single-contrast MRI inputs: MRI1 (fat-suppressed), MRI2 (water and fat), and MRI3 (water suppressed), using U-Net and our proposed architectures are shown in Figure \ref{fig:single_contrast_MR}. The performance is compared using the dice score (Figure \ref{fig:performance_parameters}). The box-plots compare U-Net and tiramisu architectures in terms of single-contrast and multi-contrast inputs. A comparison of U-Net based segmentation with our proposed methods using multi-contrast input for fat, bone, bone marrow, and muscle are shown in Figure \ref{fig:performance_parameters_all}. It is observed that our proposed tiramisu method with semi-supervision performs better in terms of dice. 
The dice scores for all tissues including IMAT are shown in Figure \ref{fig:ssdice}, when semi-supervised method was used. We observed small variance in the dice scores for all tissues except IMAT, which shows the robustness of our proposed semi-supervised method.

\subsection{Faster Convergence with Improved Dropouts}
There was no significant difference (in terms of accuracy) in using different dropout mechanisms. However, VD and TD demonstrated faster convergence compared to regular dropout. For regular dropout, we observed convergence after $800$ epochs, whereas for VD and TD, $200$ epochs were required. The convergence process for VD and TD was smoother as compared to regular dropout. Similar patterns were observed for all tissues including muscle, fat, bone, bone marrow, and IMAT. We also observed this significant improvement for all folds. Overall, the results for VD were superior to the results for both regular and TD, and therefore VD was chosen for our semi-supervised method. The segmentation results with the semi-supervised method using VD reveals a significant improvement over the state-of-the-art with mean dice scores of $97.52\%$, $94.61\%$, $80.14\%$, $95.93\%$, and $96.83\%$ for muscle, fat, IMAT, bone, and bone marrow tissues, respectively. 

\subsection{Performance Comparison}
We compared the proposed strategy with the state-of-the-art methods (as summarized in Table \ref{tab:comp}) and showed that our semi-supervised deep learning algorithm outperformed them by a significant margin, especially with IMAT segmentation. It is important that a single model was able to segment all five tissues with such high performance. A deep learning framework was used for a small cohort of subjects for generating segmentation maps of fat and muscle tissue using thigh MRI scans~\cite{bocchieri2019multiparametric}. The results were reported separately for healthy subjects and patients suffering from limb girdle dystrophy. The dice scores for both fat and muscle tissue are significantly lower as compared to our proposed methodology. A fuzzy connectivity based method (previous state-of-the-art method) was proposed for thigh tissue segmentation using MRI scans \cite{irmakci2018novel}, where the same dataset (from BLSA) was used for evaluating the system. In comparison, our proposed method outperforms the fuzzy connectivity based method in dice score by a large margin for both muscle and fat segmentation. IMAT segmentation was not addressed in this study. For muscle and IMAT, our proposed method clearly outperforms \cite{imat_perf}, which adopted a holistic neural network approach for muscle and IMAT segmentation. The results are particularly promising for IMAT, which is more challenging in terms of segmentation. These results will be used in future to detect fat infiltration within the muscle, which is of significance in aging and muscular atrophy studies. 

\begin{table}[!t]
    \centering
    \caption{Comparison with state-of-the-art methods. DSC = dice score}\label{tab:comp}
    \begin{tabular}{c|c|c}
    \toprule
    \hline
    \textbf{Method} & \textbf{DSC}  & \textbf{Tissue}\\
        
    \midrule
       Bocchieri et. al. (healthy) \cite{bocchieri2019multiparametric} & $92.00$ & \multirow{4}{*}{Fat}\\
       Bocchieri et. al. (patients) \cite{bocchieri2019multiparametric} & $88.00$ & \\
       Irmakci et al.~\cite{irmakci2018novel}  &  $84.04$  &  \\
       
       \textbf{\thead{Proposed semi-supervised}} & \textbf{94.61}  &\\
       
       \midrule
        Bocchieri et. al. (healthy) \cite{bocchieri2019multiparametric} & $88.00$ & \multirow{5}{*}{Muscle}\\
       Bocchieri et. al. (patients) \cite{bocchieri2019multiparametric} & $82.00$ & \\
        Irmakci et al.~\cite{irmakci2018novel}  & $87.18$ & \\
        
        Yao et al.~\cite{imat_perf}  & $96.90$  &\\
        
        \textbf{\thead{Proposed semi-supervised}} & \textbf{97.52}  &\\ \hline
        
        
        Yao et al.~\cite{imat_perf}  & $74.80$  & \multirow{2}{*}{IMAT}\\
        \textbf{\thead{Proposed semi-supervised}} & \textbf{80.14}  &\\
        
       \bottomrule
    \end{tabular}
\end{table}

\section{Conclusion}
To the best of our knowledge, previous studies have not dealt with addressing all tissues in the thigh region with deep learning. Also, until now, semi-supervised deep learning and multi-contrast MRI scans have not been combined to solve the thigh tissue segmentation problem yet. In this regard, our proposed semi-supervised deep learning based segmentation strategy is novel and its success has been demonstrated with a comprehensive set of experiments. In our study, we also revisited the dropout concept in deep learning, and showed the use of targeted and variational dropout functions for faster convergence and more robust segmentation. These gains were possible due to better objective function and statistical weight selection in dropout targets. We empirically observed that such inference-based dropout methods were better suited for challenging segmentation tasks when labeled data was scarce. We successfully segmented five tissues in thigh MRIs with our proposed end-to-end semi-supervised approach giving state-of-the-art results. A notable performance was achieved for IMAT (dice score: $80.14\%$, sensitivity: $88.15\%$, specificity: $99.44\%$), which is a very challenging tissue to segment. Our future studies will include exploration of the underlying theory behind statistical selection strategies for regularization and semi-supervised deep learning.

\section{Author contributions}
 UB  is the PI of the project as a part of the larger effort for creating Musculoskeletal AI in multiple institutes, agreed by the participating institutes' PIs: MA, JE, CA, SJ, and DAT.  SMA structured the overall experiments and discussions while II, SJ, and SMA ran the image analysis pipelines and deep learning experiments. UB, SJ, DAT, JE, MA, GZP, and SMA analyzed the results, and evaluated the clinical relevance, reproducibility, and feasibility of the proposed method(s). DAT and JE participated to the study as radiologists while CA and SJ contributed as MRI physicists. All authors wrote and edited the manuscript, and agreed on the content prior to submission. 

\section{Competing interests}
The authors declare no competing interests.

\section{Data Availability}
The data sets generated during and/or analyzed during the current study are available from the BLSA with research agreement~\cite{10.3945/ajcn.2009.28047}.

\section{Acknowledgment}
We would like to acknowledge the Baltimore Longitudinal Aging Study (BLAS) for providing the data set used in this study. We would also like to thank NVIDIA for donating a Titan X GPU for our experiments.  





%
%
%
%
%
\bibliographystyle{spphys}
\bibliography{ref}

\end{document}